\documentclass[article]{aa} 
\usepackage{graphicx}
\usepackage{txfonts}
\usepackage[breaklinks=true]{hyperref} 
\usepackage{natbib,twoopt}
\usepackage{float}
\usepackage{arydshln}
\usepackage[breaklinks=true]{hyperref} 
\bibpunct{(}{)}{;}{a}{}{,}    
\newcommand{\kms}{\mbox{${\rm km\,s}^{-1}$}}
\newcommand{\mjyb}{\mbox{${\rm mJy\,beam}^{-1}$}}
\newcommand{\MJup}{\ensuremath{M_{\mathrm{Jup}}}\xspace}
\newcommand{\MEarth}{\ensuremath{M_{\mathrm{Earth}}}\xspace}

\newcommand{\MSun}{\ensuremath{M_{\odot}}\xspace}

\newcommand{\LSun}{\ensuremath{L_{\odot}}\xspace}
\newcommand{\Teff}{\ensuremath{T_{\rm eff}}\xspace}
\newcommand{\Ha}{\ensuremath{\mathrm{H}\alpha}}
\newcommand{\ew}{\ensuremath{\mathrm{EW}($\Ha$)}\xspace}
\newcommand{\fw}{\ensuremath{\mathrm{FW0.1}($\Ha$)}\xspace}

\newcommand{\vsin}{\ensuremath{v \sin i}\xspace}
\newcommand{\mic}{$\,\mu$m\xspace}

\begin{document} 

\title{DZ Cha: a bona fide photoevaporating disc
    \thanks{Based on observations collected at the European Organisation for Astronomical Research in the Southern Hemisphere under ESO programme 097.C-0536.}
    \thanks{Based on data obtained from the ESO Science Archive Facility under request number $250112$.}
}

\author{H. Canovas\inst{1,2}, 
   B. Montesinos\inst{3},
   M.~R. Schreiber\inst{4,11,12},
   L.~A. Cieza\inst{5,11},
   C. Eiroa\inst{1},
   G. Meeus\inst{1},
   J. de Boer\inst{6},
   F. M\'enard\inst{7},
   Z. Wahhaj\inst{8},
   P. Riviere-Marichalar\inst{9},
   J. Olofsson\inst{4,12},
   A. Garufi\inst{1},
   I. Rebollido\inst{1},
   R. G. van Holstein\inst{6},
   C. Caceres\inst{10, 11, 12},
   A. Hardy\inst{4, 11},
   E. Villaver\inst{1}
}

\institute{ 
Departamento de F\'isica Te\'orica, Universidad Aut\'onoma de Madrid, Cantoblanco, 28049, Madrid, Spain.\\
\email{hcanovas@sciops.esa.int} 
\and European Space Astronomy Centre (ESA), Camino Bajo del Castillo s/n, 28692, Villanueva de la Ca\~nada, Madrid, Spain.
\and Dept. of Astrophysics, Centre for Astrobiology (CAB, CSIC-INTA), ESAC Campus, Camino Bajo del Castillo s/n, 28692 Villanueva de la Ca\~nada, Madrid, Spain.
\and Departamento de F\'isica y Astronom\'ia, Universidad de Valpara\'iso, Valpara\'iso, Chile.
\and Facultad de Ingenier\'ia y Ciencias, Universidad Diego Portales, Av. Ejercito 441, Santiago, Chile.
\and Leiden Observatory, Leiden University, P.O. Box 9513,2300RA Leiden, The Netherlands.
\and Univ. Grenoble Alpes, CNRS, IPAG, F-38000 Grenoble, France.
\and European Southern Observatory, 3107 Alonso de Cordova, Vitacura, 1058 Santiago, Chile.
\and Instituto de Ciencia de Materiales de Madrid (ICMM-CSIC). E-28049, Cantoblanco, Madrid, Spain.
\and Departamento de Ciencias Fisicas, Facultad de Ciencias Exactas, Universidad Andres Bello. Av. Fernandez Concha 700, Las Condes, Santiago, Chile.
\and Millennium Nucleus ``Protoplanetary discs in ALMA Early Science".
\and Millennium Nucleus for Planet Formation.
}

\date{\today}

\abstract
{DZ Cha is a weak-lined T Tauri star (WTTS) surrounded by a bright protoplanetary disc with evidence of inner disc clearing. Its
narrow $\Ha$ line and infrared spectral energy distribution suggest that DZ Cha may be a photoevaporating disc.}
{We aim to analyse the DZ Cha star + disc system to identify the mechanism driving the evolution of this object. }
{We have analysed three epochs of high resolution optical spectroscopy, photometry from the UV up to the sub-mm regime, infrared
spectroscopy, and J-band imaging polarimetry observations of DZ Cha.}
{Combining our analysis with previous studies we find no signatures of accretion in the $\Ha$ line profile in nine epochs covering a time
baseline of $\sim20$ years. The optical spectra are dominated by chromospheric emission lines, but they also show emission
from the forbidden lines [SII] 4068 and [OI] 6300$\,\AA$ that indicate a disc outflow. The polarized images reveal a dust depleted cavity of $\sim7$ au
in radius and  two spiral-like features, and we derive a disc dust mass limit of $M_\mathrm{dust}<3\MEarth$ from the sub-mm photometry. No stellar
($M_\star > 80 \MJup$) companions are detected down to $0\farcs07$ ($\sim 8$ au, projected).}
{The negligible accretion rate, small cavity, and forbidden line emission strongly suggests that DZ Cha is currently at the initial stages of disc clearing by photoevaporation.
At this point the inner disc has drained and the inner wall of the truncated outer disc is directly exposed to the stellar radiation. We argue that other mechanisms like planet
formation or binarity cannot explain the observed properties of DZ Cha. The scarcity of objects like this one is in line with the dispersal timescale ($\lesssim 10^5$ yr)
predicted by this theory. DZ Cha is therefore an ideal target to study the initial stages of photoevaporation.}


\titlerunning{DZ Cha: a bona fide photoevaporating disc}
\authorrunning{Canovas et al.}

\maketitle

\section{Introduction}
\label{sec:intro}

T Tauri stars (TTS) are pre-main sequence stars (PMS) of late spectral types (SpT, F-M) and typical masses $< 1.5\MSun$
\citep[e.g. ][]{1989ARA&amp;A..27..351B}.  They are surrounded by circumstellar discs created during the collapse of the
protostellar cores. Since their discovery by \cite{1945ApJ...102..168J}, TTS have attracted wide attention from the astronomy
community, and nowadays it is accepted that planets can form in their surrounding discs \citep{2015Natur.527..342S, 2015ApJ...807...64Q}.
Understanding how these discs evolve is therefore key to constraining theories of planet formation and  eventually to explaining
the diversity of currently observed planetary architectures  \citep[e.g.][and references therein]{2015ARA&amp;A..53..409W}.

Historically, TTS have been classified into two groups according to their accretion tracers (mainly the
$\Ha$ line equivalent width, $\ew$). Classical T Tauri stars (CTTS) show broad $\Ha$ line profiles
(usually  $\ew \geq 10\,\AA$), strong emission lines at ultraviolet (UV) and optical wavelengths, and/or veiling
signatures across the UV and optical ranges. These observables indicate active accretion of material onto the star, with
mass accretion rates ranging from $10^{-7} - 10^{-9}\, M_{\sun}$ yr$^{-1}$ \citep{1998ApJ...492..323G}.
Thermal emission from their dusty circumstellar discs is usually detected above photospheric levels in nearly the entire
infrared and sub-mm regime \citep[e.g. ][]{2009ApJS..181..321E, 2013ApJ...776...21H, 2016ApJ...828...46A}. On the
other hand, Weak-lined T Tauri stars (WTTS)  show no evidence of accretion activity. Their spectra show
no veiling signatures, and they lack most of the emission lines observed in CTTS. Their $\Ha$ line is mostly dominated
by chromospheric emission and is typically narrow ($\ew < 10\,\AA$). Hot circumstellar dust, probed at near- and
mid-infrared wavelengths, is detected in only  $\sim 20\%$ of them \citep{2006ApJ...645.1283P, 2007ApJ...667..308C,
2010ApJ...724..835W}, and this excess is systematically fainter than in CTTS. In the sub-mm regime their discs are
usually fainter than the more luminous discs around CTTS \citep{2005ApJ...631.1134A}, and some of them have
disc masses comparable to the gas-poor debris discs \citep[][]{2015A&amp;A...583A..66H}.
Both CTTS and WTTS are bright X-ray sources, and in fact WTTS (originally called  \textit{naked} TTS)
were first discovered via X-rays surveys \citep[][]{1988AJ.....96..297W}.

As they evolve, all CTTS will become WTTS before entering the main sequence (MS). In other words, the discs around
CTTS will eventually be very depleted of gas and dust, and accretion onto the star will be too faint to leave a detectable imprint in
the stellar spectrum. Discs showing evidence of inner cavities and/or gaps are thought to be in transition between these two evolutionary
stages \citep{1989AJ.....97.1451S}. The scarcity of these objects indicates that this transition phase lasts $\lesssim0.5$ Myr
\citep[e.g. ][]{1996AJ....111.2066W, 2007ApJ...667..308C}, and multiwavelength observations indicate that once the process
begins the entire disc quickly dissipates from inside out \citep{2000A&amp;A...355..165D}. Photoevaporation from the central star
in combination with viscous accretion is, to date, the most successful theory in explaining this rapid transition \citep{2014prpl.conf..475A}.
Current observational evidence supports this theory;  several discs show photoevaporative winds traced by the
[O I] ($6300\AA$) and [Ne II] ($12.8$\mic) forbidden emission lines and by free-free emission at centimetre wavelengths
\citep{2009ApJ...702..724P, 2012ApJ...747..142S, 2013ApJ...772...60R, 2016ApJ...829....1M}.
Photoevaporation models show that beyond a critical radius (a few au for $1\, \MSun$ stars), the stellar radiation creates a pressure
gradient in the disc upper layers that drives a photoevaporative wind  \citep{1994ApJ...428..654H, 2004ApJ...607..890F}. 
This process strongly depends on the energy regime dominating the photoevaporation (far-ultraviolet, FUV; extreme
ultraviolet, EUV; and/or X-rays), with predicted photoevaporation rates ranging from $10^{-8} - 10^{-10}\, M_{\sun}$ yr$^{-1}$
\citep[][]{2009ApJ...699.1639E, 2009ApJ...690.1539G, 2012MNRAS.422.1880O}. These models
qualitatively agree on the initial steps of disc clearing: once the stellar mass accretion rate drops to the level of the photoevaporation mass
loss rates, the flow of material from the outer disc towards the star is interrupted at the critical radius. When this happens the inner disc
(the region encompassed by the critical radius) is detached from the outer disc and it is quickly accreted by the star \citep{2001MNRAS.328..485C}.
At this point, the inner edge of the outer disc is directly exposed to the stellar radiation and the entire outer disc disperses from the inside out in
$\sim10^5$ yr \citep{2006MNRAS.369..216A, 2006MNRAS.369..229A}. Despite the success in explaining the evolution of protoplanetary discs,
observations confirming the first steps of the photoevaporation predicted by the theory are still missing.

DZ Cha (alias PDS 59; RX J1149.8-7850, $\alpha = 11^\mathrm{h}49^\mathrm{m}31.84^\mathrm{s}$, $\delta = -78\degr51\arcmin01.1\arcsec$)
is a peculiar M0Ve TTs \citep{1992AJ....103..549G, 2006A&amp;A...460..695T}. Although studies link it to the nearby $\beta$ Pic association
\citep[][]{2006AJ....132..866R, 2013ApJ...762...88M, 2014ApJ...788...81M}, given its proper motions and Li 6708$\,\AA$ equivalent width
DZ Cha most likely belongs to the more distant $\epsilon$ Cha sparse stellar association \citep[][]{2006A&amp;A...460..695T, 2008hsf2.book..757T,
2008ApJ...675.1375L, 2013A&amp;A...551A..46L, 2013MNRAS.435.1325M, 2014A&amp;A...568A..26E}. In a detailed study of the $\epsilon$ Cha
association, \cite{2013MNRAS.435.1325M} derive a distance of $d = 110\pm7$ pc and an age of 2-3 Myr for DZ Cha. This star shows strong flaring events
\citep{1997A&amp;A...321..803G, 2000MmSAI..71.1021T}, and  was first classified as a WTTS by \cite{1995A&amp;AS..114..109A}. 
DZ Cha has a very narrow $\Ha$ line profile, with a full width at $10\%$ of the line of $\fw \sim 120\,\kms$ and equivalent width
$\ew < 5\,\AA$ \citep{2006A&amp;A...460..695T, 2010ApJ...724..835W}. Its near-infrared (NIR) SED shows evidence of an inner cavity,
but at longer wavelengths its strong excess is virtually indistinguishable from the SEDs of accreting discs around CTTS 
\citep{2008ApJ...675.1375L, 2010ApJ...724..835W}. In fact, its fractional disc luminosity is the highest among the hundreds of WTTS
observed during the Spitzer `c2d'  program \citep{2010ApJ...724..835W}.

Given the properties of DZ Cha we selected this object as an ideal target to study the transition from CTTS to WTTS. In this paper we present
the first detailed analysis of this interesting disc. We show that the evidence of inner disc clearing in combination with its negligible accretion
rate across different epochs, strong disc emission at  mid-infrared wavelengths, and the disc outflow probed by forbidden emission lines, 
are best explained if DZ Cha is at the initial stages of inner disc clearing by photoevaporation. Our observations and results are presented
in Sects.~\ref{sec:obs} and \ref{sec:results}, respectively. The results are discussed in Sect.~\ref{sec:discusion}, and a summary and
conclusions are presented in Sect.~\ref{sec:conclusions}.

\section{Observations and data reduction}
\label{sec:obs}
We have combined three epochs of high resolution optical spectroscopy,  photometry ranging from the UV to the
sub-mm regime, and imaging polarimetry at J band. Below we describe the observations and data processing.

\subsection{Optical spectroscopy}
\label{subsec:spec}

High resolution (processed) spectroscopy at optical wavelengths was retrieved from the ESO Science Archive facility. The observations
were acquired with the Ultraviolet and Visual Echelle Spectrograph (UVES) at the Very Large Telescopes (VLT's) as part of the ESO
program ID 088.C-0506 (PI: C. Torres) over three nights. A summary of the observations is given in Table~\ref{tab:tab_uves_log}.
The instrument was configured using the $1\farcs0$ slit that samples the $3250-6800\, \AA$ range with a resolution of
$\lambda / \Delta \lambda \sim 40 000$. Each of the three datasets has a signal-to-noise ratio (S/N) of $\sim 80$ in the continuum
around the Li 6708$\,\AA$ line. Barycentric and stellar radial velocity \citep{2014A&amp;A...568A..26E} corrections were
applied with customized scripts.
\begin{table}[h!]
\center
\caption{UVES observation log.}
\begin{tabular}{cccc}
\hline
Date              & <Seeing>       & <Airmass> & Int. Time\\
(yy-mm-dd)   & ($\arcsec$)       &                   & (s)\\

\hline \hline
2012-01-09 & $0.80$   & 1.77 & 575\\
2012-02-24 & $0.94$ & 1.72 & 575\\
2012-03-07 & $0.91$ & 1.74 & 575\\
\hline
\end{tabular}
\tablefoot{The columns list the observation date,  average airmass, seeing, and total integration time.}
\label{tab:tab_uves_log}
\end{table}

\subsection{Photometry}
\label{subsec:phot}

In this work we  use calibrated photometry publicly available in the  Vizier Service\footnote{http://vizier.u-strasbg.fr/viz-bin/VizieR} and
at the NASA/IPAC Infrared Science Archive (IRSA)\footnote{http://irsa.ipac.caltech.edu/frontpage/}.
Ultraviolet (UV) photometry was obtained from the GALEX catalogue of UV sources \citep{2011Ap&amp;SS.335..161B}. Optical photometry
(Johnson B and  V, and Sloan g', r', and i') was obtained from the American Association of Variable Star Observers (AAVSO) Photometric All
Sky Survey \citep[APASS, ][]{2016yCat.2336....0H}. 
Infrared photometry was obtained from the Two Micron All Sky Survey \citep[2MASS, ][]{2006AJ....131.1163S},
from the Spitzer c2d legacy program \citep{2003PASP..115..965E}, from the ALLWISE catalogue \citep{2014yCat.2328....0C},
and from the Herschel/SPIRE point source catalogue. We also obtained the Spitzer/IRS spectra acquired during the Spitzer GO
program \#50053 (PI: Houck, J. R.). Visual inspection of these catalogues shows no companions and/or background emission that
may contaminate the published photometry and IRS spectra.

Furthermore, we retrieved raw Herschel/PACS (70 and 160 $\mu$m) observations from the Herschel Science Archive\footnote{http://archives.esac.esa.int/hsa/whsa/}
and processed them following \cite{2013A&amp;A...555A..67R}. The
Herschel Interactive Processing Environment (HIPE) 14 with the most recent version for the calibration files was used to identify and remove
bad pixels, to apply flat field correction, for deglitching, high pass filtering, and map projection. Aperture photometry was measured in the processed
images using an aperture of $6\arcsec$ and $12\arcsec$ for the 70 and 160$\,\mu$m images, respectively. The sky was measured in a ring
 centred on the star with radius and width of $25\arcsec$ and $10\arcsec$, respectively. The final uncertainties in the photometry were computed
as the quadratic sum of the photometry and calibration errors.

Sub-mm photometry was obtained with the Atacama Pathfinder Experiment (APEX\footnote{This
publication is based on data acquired with  APEX which is a collaboration between the Max-Planck-Institut
fur Radioastronomie, the European Southern Observatory, and the Onsala Space Observatory.}), 
the 12m radio telescope located on Llano de Chajnantor (Chile). Our observations were performed
during period 092 (C-092.F-9701A-2013). We used the APEX-LABOCA camera \citep{2009A&amp;A...497..945S}
which operates at a central frequency of  345 GHz ($870\,\mu$m) aiming to detect the dust continuum
emission from DZ Cha. The APEX pointing uncertainty is $2$'' \citep{2006A&amp;A...454L..13G}, and
the nominal LABOCA beam full width at half maximum (FWHM) is $19.2\pm0.7$''. The most sensitive
part of the bolometer array was centred on the target to maximize the chance of detection. Skydips and
radiometer measurements were interleaved during the observation to obtain accurate atmospheric opacity
estimates. Mars and the quasar PKS1057-79 were used to calibrate the focus and pointing, respectively.
The secondary calibrator NGC 2071 was used to perform the absolute flux calibration, which is expected
to be accurate to within 10$\%$ \citep[][]{2009A&amp;A...497..945S}.
The emission around DZ Cha was mapped following a raster map in spiral mode. The weather conditions
were excellent with a nearly constant precipitable water vapour (PWV) of $0.22$~mm. A total of ten scans
were obtained, resulting in 1.52 h of on-source integration time. The observations were reduced with
the CRUSH package \citep{2008SPIE.7020E..1SK} using the \textit{faint} reduction mode and half a beam
smoothing factor. This way we reached an rms of 8 $\mjyb$, but no emission was detected at the location of DZ Cha. 

The observed and de-reddened fluxes (see Sect.~\ref{subsubsec:extinction}) are listed in Table~\ref{tab:tab_sed}.
\begin{table}[t]
\center
\caption{SED of DZ Cha. The columns list the effective wavelength of each filter, the observed flux, the de-reddened flux, and the associated errors.
Extinction uncertainties dominate at wavelengths shorter than 1$\,\mu$m, while calibration uncertainties dominate at longer wavelengths.}
\begin{tabular}{ccccc}
\hline
Wavelength & Flux & Flux Dered. & Error & Ref. \\
$(\mathrm{\mu m})$ & $(\mathrm{mJy})$ & $(\mathrm{mJy})$ & $(\%)$ &  \\
\hline \hline
0.15 & 2.20E-02 & 5.46E-02 & $30$ & 1 \\
0.23 & 8.50E-02 & 2.26E-01 & $30$ & 1 \\
0.44 & 7.90E+00 & 1.25E+01 & $15$ & 2 \\
0.48 & 1.38E+01 & 2.10E+01 & $15$ & 2 \\
0.55 & 2.85E+01 & 4.01E+01 & $10$ & 2 \\
0.62 & 5.09E+01 & 6.79E+01 & $10$ & 2 \\
0.76 & 1.17E+02 & 1.45E+02 & $10$ & 2 \\
1.24 & 2.62E+02 & 2.90E+02 & $5$ & 3 \\
1.65 & 3.41E+02 & 3.63E+02 & $5$ & 3 \\
2.16 & 2.72E+02 & 2.83E+02 & $5$ & 3 \\
3.35 & 1.66E+02 & 1.69E+02 & $5$ & 4 \\
3.55 & 1.65E+02 & 1.69E+02 & $5$ & 5 \\
4.49 & 1.73E+02 & 1.76E+02 & $5$ & 5 \\
4.60 & 1.54E+02 & 1.57E+02 & $5$ & 4 \\
5.73 & 1.90E+02 & 1.93E+02 & $5$ & 5 \\
7.87 & 3.66E+02 & 3.72E+02 & $5$ & 5 \\
11.56 & 4.42E+02 & 4.44E+02 & $5$ & 4 \\
22.09 & 1.54E+03 & 1.55E+03 & $5$ & 4 \\
23.67 & 1.24E+03 & 1.24E+03 & $10$ & 5 \\
70.00 & 1.73E+03 & 1.73E+03 & $5$ & 6 \\
160.00 & 1.09E+03 & 1.09E+03 & $5$ & 6 \\
250.00 & 3.48E+02 & 3.48E+02 & $5$ & 6 \\
363.00 & 1.59E+02 & 1.59E+02 & $5$ & 6 \\
517.00 & 7.41E+01 & 7.41E+01 & $15$ & 6 \\
870.00 & <2.40E+01$^{\mathrm{a}}$ & <2.40E+01$^{\mathrm{a}}$ & ... & 6 \\
\hline
\end{tabular}
\tablefoot{$^{\mathrm{a}}$: $3\sigma$ upper limit.
Photometry references: 
1) \cite{2011Ap&amp;SS.335..161B};
2) \citet{2016yCat.2336....0H}; 
3) \citet{2003yCat.2246....0C}; 
4) \citet{2014yCat.2328....0C}; 
5) \citet{2009ApJS..181..321E};
6) This work.}
\label{tab:tab_sed}
\end{table}

\subsection{Imaging polarimetry}
\label{subsec:pol_img}

We observed DZ Cha with the SPHERE \citep[Spectro-Polarimetric High-contrast Exoplanet REsearch, ][]{2008SPIE.7014E..18B}
instrument at the VLT on April 4, 2016, under ESO programme 097.C-0536(A). Observations were carried out with the infrared
subsystem IRDIS \citep[Infra-Red Dual Imaging and Spectrograph,][]{2008SPIE.7014E..3LD} in the dual polarization imaging mode
\citep[DPI]{2014SPIE.9147E..1RL} with and without coronagraph mask in the broad-band J filter of SPHERE ($\lambda_c = 1.245\,\mu$m).
Both sets were divided in polarimetric cycles where each cycle contains four datacubes, one per half-wave plate (HWP) position
angle (at $0\degr, 22.5\degr, 45\degr$, and $67.5\degr$, measured on sky east from north). The complete observation sequence amounted
to 1.9 h on-source at airmass ranging from 1.7 to 1.8, and standard calibration files (darks and flat fields) were provided by the ESO observatory.

The coronagraphic observations were taken with an apodized pupil Lyot coronagraph with a focal mask of $\sim 0\farcs09$ in radius
\citep{2011ExA....30...39C}. Centre calibration frames with four satellite spots produced by a waffle pattern applied to the deformable mirror
were taken at the beginning and at the end of the coronagraphic sequence to determine the position of the star behind the mask. 
In total, the coronagraphic observations amounted to  1.6 h on-source, with a median seeing of $0\farcs74 \pm 0\farcs12$ and coherence
time of $5\pm1$ ms. The non-coronagraphic observations amounted to 0.3 h on-source under better weather conditions, with a median
seeing of $0\farcs49 \pm 0\farcs06$ and coherence time of $8\pm1$ ms. We used detector integration times (DITs) of 64s and 2s for
the coronagraphic and non-coronagraphic observations, respectively. A summary of the observations is given in Table~\ref{tab:tab_sphere_log}.

The  HWP projects two simultaneous images with orthogonal polarization states over different regions on the detector. Subtracting these
two images when the HWP is at $0\degr (45\degr)$ yields the Stokes parameter $Q^+(Q^-)$. Repeating this process for the $22.5\degr (67.5\degr)$
angles produces the Stokes $U^+ (U^-)$ images. The total intensity (Stokes I) is computed by adding  (instead of subtracting) the two
images delivered by the HWP. We used customized scripts to process the raw data following the high-contrast imaging polarimetry
pipeline described by \citet{2011A&A...531A.102C}. First, each science frame was dark current subtracted and flat-field corrected.
Bad pixels (hot and dead)  were identified with a $\sigma$ clipping algorithm and masked out using the average of their surrounding
good pixels. The two simultaneous images from each science frame were first aligned using a cross-correlation algorithm. After this we
applied an algorithm that finds the alignment that minimizes the standard deviation of the difference between the two images by shifting
each image in steps of 0.05 pixels \citep[see also][]{2015A&amp;A...578L...1C}. This process was applied to every science frame resulting
in a datacube for each Stokes $Q^\pm, U^\pm$ parameter. These images were combined using the double-difference method \citep{2011A&A...531A.102C},
yielding the observed Stokes $Q$ and $U$ parameters. At this stage we correct for instrumental polarization and instrument-induced cross-talk
with the Mueller matrix model of \cite{2017arXiv170907519V} and de Boer et al. (in prep.). This model describes the complete telescope+instrument
system and is based on measurements with SPHERE's internal calibration lamp and observations of an unpolarized standard star.
In this correction method every measurement of Stokes $Q$ and $U$
is described as a linear combination of the Stokes $I$-, $Q$-, and $U$-images incident on the telescope. The incident $Q$- and $U$-images
were obtained by solving, using linear least-squares, the system of equations describing all measurements. We then derived the polarization
angle ($P_{\theta}=  0.5\, \arctan \left(U/Q\right)$), the polarized intensity ($P_{I} = \sqrt{Q^2 + U^2}$), and the $Q_{\phi}$ and $U_{\phi}$ images
\citep[see][]{2006A&amp;A...452..657S, 2014ApJ...781...87A}, with an accuracy below $\sim0.5^\circ$ in $P_{\theta}$. This method
is advantageous as no assumptions are made about the polarization of the star and the angle of linear polarization of the disc, and
it has  already been benchmarked against other pipelines \citep{2017arXiv170503477P}.

While the polarized images probe the disc surface, the intensity images alone can be used to derive detection limits on possible companions.
Our observations were taken in field-tracking mode and no comparison star was observed, so high contrast algorithms 
\citep[e.g. principal component analysis, ][]{2012ApJ...755L..28S} cannot be applied. Instead, we applied two image filters to the individual
exposures to remove noise contributions from broad spatial features and to isolate point sources. The first filter removes azimuthally symmetric
features like the stellar halo, as described in \cite{2013ApJ...779...80W}. The second filter is similar to the first, but it removes running flux
averages over 20 pixels along the radial direction (instead of  azimuthal)  with the star as centre \citep[see][]{2016A&A...596L...4W}. This
removes radial PSF features like diffraction spikes. We then computed the noise level of the individual intensity images in concentric rings
centred over the star location.  A contrast curve was computed by determining the $5\sigma$ detection level at each radii after stacking all
the intensity images.

\begin{table}[]
\center
\caption{SPHERE observation log.}
\begin{tabular}{cccccc}
\hline
Dataset                & Seeing          & $\tau$  & DIT   & Cycles & Int. Time \\
                             &   ($\arcsec$)   &   (ms)   &  (s)    &             & (s) \\
\hline \hline
Coronagraphic      & $0.74$        & 5           & 64     & 11        & 5632\\
Non-Coro.             & $0.49$        & 8          & 2      & 14        & 1120\\
\hline
\end{tabular}
\tablefoot{The columns list the dataset, median seeing, median $\tau$, DIT, polarization cycles, and total integration time.}
\label{tab:tab_sphere_log}
\end{table}

\section{Results}
\label{sec:results}

\subsection{The star}
\label{sub_sect:stellar}

Owing to the spread in distances and stellar parameters  found in the literature, here we review the
stellar properties of DZ Cha using our observations. 

\subsubsection{Extinction}
\label{subsubsec:extinction}

Previous studies have used the intrinsic colours for main sequence (MS) M0 stars \citep{1995ApJS..101..117K} obtaining a total
extinction of $A_V = 0.9$ \citep{2010ApJ...724..835W, 2013MNRAS.435.1325M}. As DZ Cha has not yet entered the MS stage
here we used the intrinsic colours for M0 pre-main sequence stars from \citet{2013ApJS..208....9P}. Those were subtracted
from the observed $(J-H)$, $(J - K_s)$, and $(H - K_s)$ to obtain different  colour excesses. These colours are not affected
 by veiling or by disc emission (see Sects.~\ref{sub_sect:em_lines} and ~\ref{sect:disc}). The $A_V$ was computed from
each excess assuming an extinction parameter of $R_V = 3.1$ and the \cite{1999PASP..111...63F} reddening law including the
NIR empirical corrections by \citet{2005ApJ...619..931I}. We find that $A_V$ remains roughly constant independently of
the colour index used, with an average value of $A_V = 0.4\pm0.1$. We then used this value of $A_{V}$ to de-redden the entire
photometry, and through our work we consider an extinction uncertainty of $\delta A_V = 0.1$.

\subsubsection{Stellar properties}
\label{subsubsec:stellar_props}

The temperature scale for M0 PMS stars derived by \cite{2013ApJS..208....9P} yields $\Teff= 3770\pm30$ K. 
We then compared the observed de-reddened photometry to the Phoenix/NextGen \citep{1999ApJ...512..377H, 2012RSPTA.370.2765A}
grid of synthetic stellar models in this temperature range. We chose the family of models computed for solar abundances
\citep[][]{2009ARA&amp;A..47..481A}, with solar 
metallicity and zero $\alpha-$element enhancement (as expected for young stars in the solar neighbourhood), and with surface gravity
in the range $\log g = [3.5 - 4.5]$. To directly compare the models with the observations we first normalized the synthetic flux at J band
($\lambda_\mathrm{c} = 1.2\,\mu$m) to the observed value. In this band the stellar photosphere still dominates over the disc emission
while the extinction uncertainties are lower than at optical bands. Integrating the photosphere at all
wavelengths yields a stellar luminosity of $L_{\star} = 0.6\pm0.1\, L_{\sun}$, where the distance uncertainty is the dominant source of error.
Provided with the effective temperature and stellar luminosity, we then used three different evolutionary tracks
\citep{1998A&amp;A...337..403B, 1999ApJ...512..377H, 2000A&amp;A...358..593S}
to estimate the stellar mass ($M_{\star} = 0.70 \pm 0.15 M_{\sun}$), radius ($R_{\star} = 1.7 \pm 0.2 R_{\sun}$), and age ($2.2\pm0.8$ Myr). 
The radius and stellar mass yield a surface gravity of $\log g = 3.7 \pm 0.1$ cm s$^{-2}$.
\cite{1997A&amp;A...328..187C} estimated the projected rotational velocity $\vsin$ of several WTTS including DZ Cha,  where
$v$ is the rotational velocity at the stellar equator and $i$ is the stellar inclination angle. They applied two different methods\footnote{
\cite{1997A&amp;A...328..187C} explain that their cross-correlation method (which yields $\vsin = 18\pm5\,\kms$ for DZ Cha)
is more reliable for slow rotators. To the best of our knowledge, all the following studies of DZ Cha quote only this value, ignoring the
alternative estimate of $13\pm5\,\kms$.}
to their high resolution spectra ($\lambda / \Delta \lambda \sim 20000$), obtaining $\vsin = 18\pm5\, \kms$ and $\vsin = 13\pm5\, \kms$.
The UVES observations discussed in this paper have a spectral resolution that is  nearly two times higher, so they can be used to further constrain
this value. To do so, we first synthesized a high resolution spectrum from the Phoenix/NextGen family of models using the effective temperature,
metallicity, and surface gravity previously derived. This spectrum is then reddened and re-sampled at the same wavelengths as the UVES spectra.
We then used the PyAstronomy package\footnote{https://github.com/sczesla/PyAstronomy} 
to apply different rotational profiles to the synthetic spectra, including limb-darkening corrections 
\citep[using a limb-darkening coefficient of $\epsilon = 0.6$ following][]{1986A&amp;A...165..110B}. Visual inspection of the synthesized
spectra shows that models with $\vsin =18\,\kms$ produce absorption lines that are too broad, while models with $\vsin \le 13\,\kms$, and
in particular $\vsin \sim 8\,\kms$  \cite[i.e. the lowest value within error bars estimated by][]{1997A&amp;A...328..187C}, produce a better
match to our observations.
 
\citet{1998A&AS..128..561B} and \citet{2011A&amp;A...532A..10M} derived rotational periods ($P$) of 11.7 and 8.0 days, respectively, using
multiple epochs of high precision photometry at optical wavelengths. Computing the period as $P = 2 \pi \, R_{\star} / v$, using our derived values
of $R_{\star} = 1.7\,R_{\sun}$, and assuming that the stellar inclination is the same as the disc ($43\degr\pm5\degr$, see Sect.~\ref{sect:pol_images}),
we derive $P = 3.3\pm0.3$ days for $\vsin = 18\,\kms$. This discrepancy with the period values determined from the photometry was
previously noticed by \cite{2011A&amp;A...532A..10M}. In contrast, using $\vsin =  8\,\kms$ results in $P_\mathrm{max} = 7.4\pm0.7$ days,
in better agreement with the photometric periodic variability. Therefore, we adopt $\vsin \sim 8\, \kms$ as a representative value. The stellar
properties are summarized in Table ~\ref{tab:tab_stellar_props}.
\begin{table}[h!]
\center
\caption{Stellar properties.}
\begin{tabular}{ccc}
\hline
Parameter & Symbol & Value \\
\hline
\hline
Spectral Type$^{a}$        & SpT            & M0Ve \\
Distance$^{b}$               & $d$             &$110\pm7$ pc\\
Visual Extinction              & $A_{V}$    & $0.4\pm0.1$ \\
Effective Temperature     & $\Teff$       & $3770\pm30$ K \\
Luminosity                      & $L_{\star}$ & $0.6 \pm0.1$ $L_{\sun}$ \\
Radius                            & $R_{\star}$ & $1.7 \pm0.1$ $R_{\sun}$ \\
Mass                               & $M_{\star}$ & $0.7 \pm0.2$ $M_{\sun}$ \\
Surface gravity               & $\log g$       & $3.7 \pm0.1$ cm s$^{-2}$ \\
Age                                 &                    & $2.2 \pm0.8$ Myr \\
Projected Velocity           &  $\vsin$       & $\sim 8\, \kms$ \\
\hline
\end{tabular}
\tablefoot{$^{a}$ Spectral type from \cite{2006A&amp;A...460..695T}; $^{b}$ distance from \cite{2013MNRAS.435.1325M}.}
\label{tab:tab_stellar_props}
\end{table}
\begin{figure*}[]
  \centering \includegraphics[width=\textwidth, trim = 0 40 0 0]{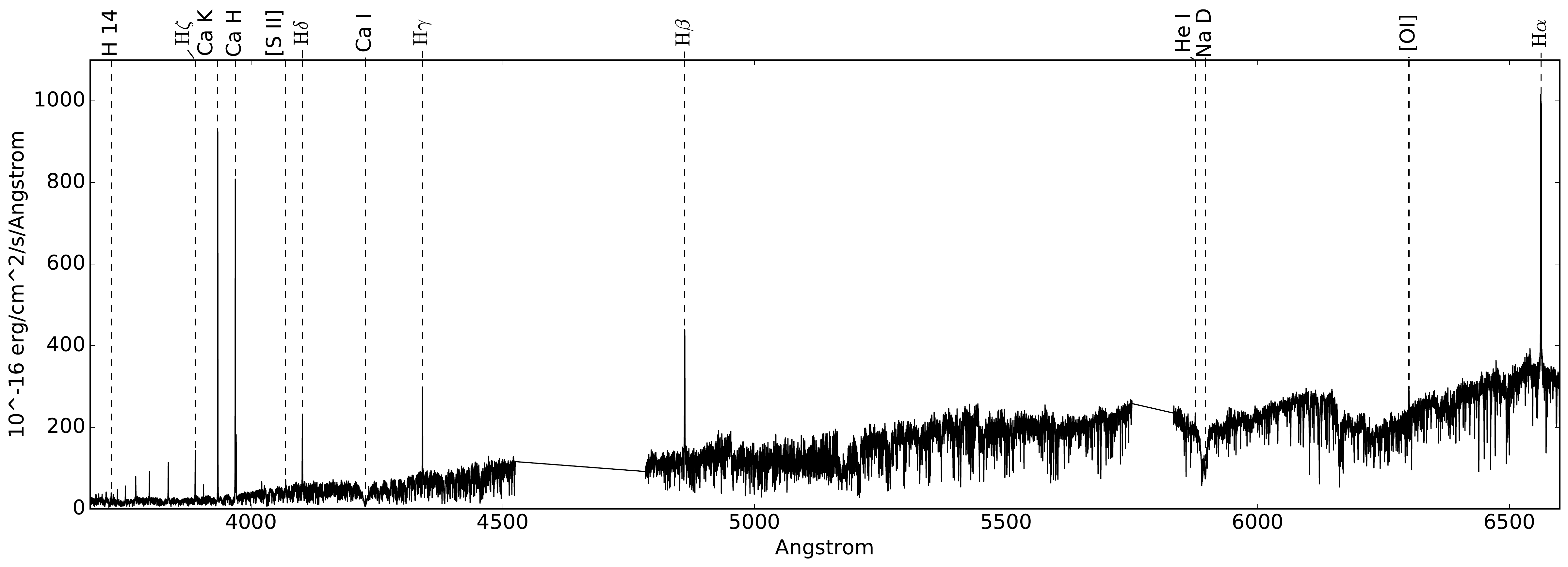}
  \caption{Entire UVES spectrum of DZ Cha with selected emission lines highlighted. The hydrogen Balmer series is detected in emission
  up to the 14-2 ($3721.946\,\AA$) transition. Only one spectra is shown for clarity, as the three different spectra analysed here show almost
  no difference.}
   \label{fig:all_uves}
\end{figure*}

\subsubsection{Emission and absorption lines}
\label{sub_sect:em_lines}

\begin{figure}[]
  \centering \includegraphics[width=0.8\columnwidth, trim = 0 20 0 0]{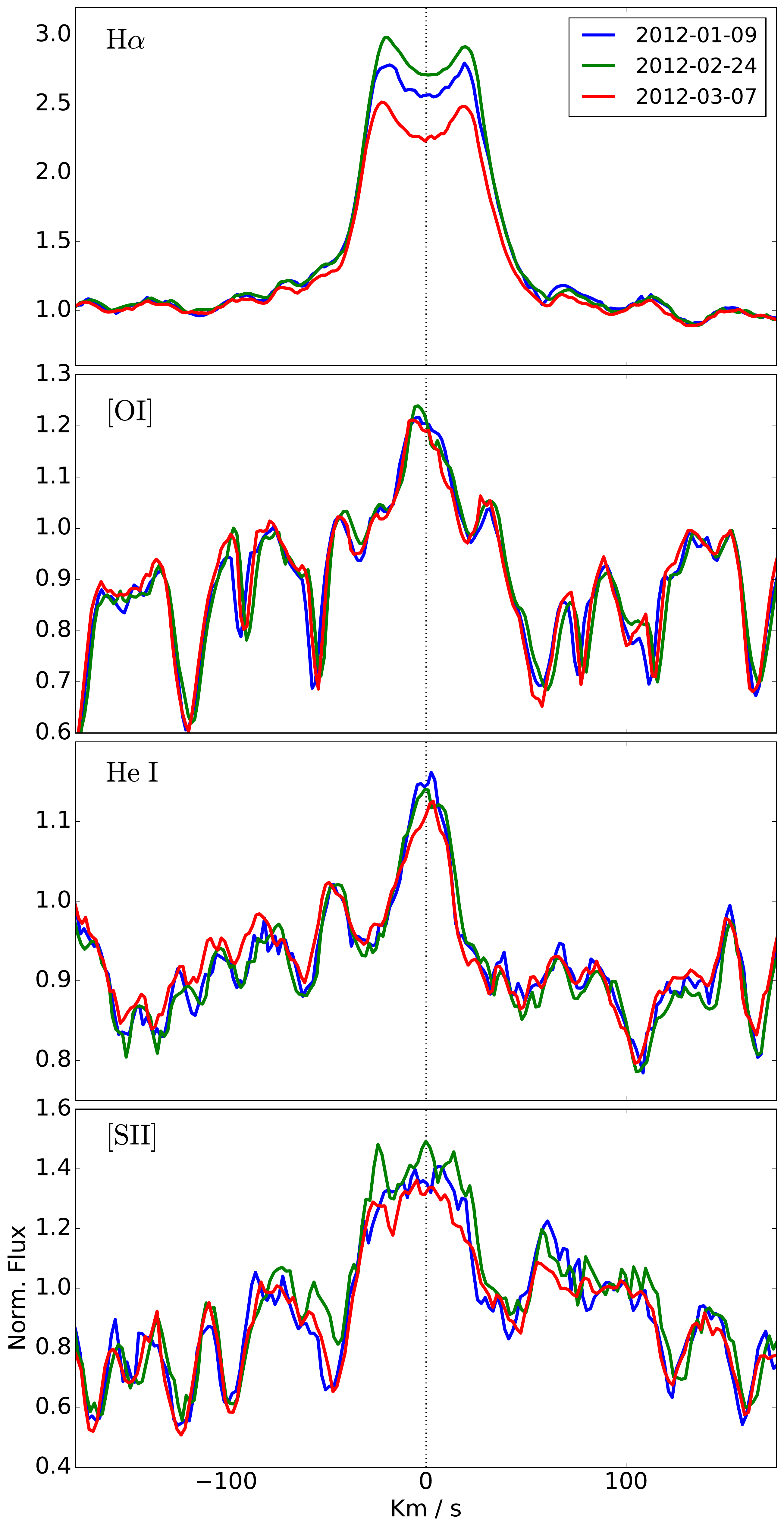}
  \caption{Continuum normalized emission lines. The [OI] line is blue-shifted in all nights, peaking at $\sim -5.1\, \kms$.}
   \label{fig:fig_em_lines}
\end{figure}

The optical spectra contain a wealth of prominent emission lines including the Balmer series up to H14, a few iron lines,
and faint but significant emission in the [OI] 6300$\,\AA$ and [S II] 4068$\,\AA$ forbidden lines (Fig.~\ref{fig:all_uves}).
Given their importance as accretion and/or stellar activity tracers, we list the equivalent widths for the $\Ha$ and $H{\beta}$
lines, the Ca K and H doublet, the He I 5876$\,\AA$\footnote{
This line is a blend of the triplet He I $^{3}$D transition.},
and the [OI] and [S II] forbidden lines in Table~\ref{tab:tab4}. The normalized profiles for $\Ha$, He I, [OI], and
[S II] are shown in Fig.~\ref{fig:fig_em_lines}. There is no significant variation in the equivalent width in any
of the lines except in $\Ha$, which shows a decrement from night 2 to night 3. We also computed the full width at $10\%$ of
the $\Ha$ line, finding that $\fw  = 128\pm2\,\kms$ during the three nights. 
We have not found calibrated relations in the literature for M0 stars with $\fw < 150\,\kms$, and using the relations derived by
\cite{2004A&amp;A...424..603N} would result in an upper limit of $\dot{M} < 10^{-11} \MSun \, \mathrm{yr}^{-1}$. However, 
low ($\dot{M} < 10^{-10} \MSun \, \mathrm{yr}^{-1}$) accretion rates derived from the $\Ha$ line profile alone can be unreliable
because of contamination by chromospheric activity \citep[e.g.][]{2011ApJ...743..105I, 2013ApJ...767..112I, 2013A&amp;A...551A.107M}.
The $\Ha$ line has a centro-symmetric double-peaked profile, very similar
(both in shape and in equivalent width) to the $\Ha$ profiles observed around gas-poor discs like AU Mic \citep{1994A&A...289..185H}. The
[OI] line is blue-shifted on all nights, with peak values ranging from -3.5 to -6.8 \kms. Fitting a Gaussian profile to the [OI] line results in
Gaussian centres located at $-2.9 \pm 0.3 $ \kms\ and full width at half maximum FWHM $\sim 46\pm4$ \kms.
Using the optical photometry and the line equivalent width we obtain a line luminosity of $L_\mathrm{[O I]} = 3.1\times10^{-6} \LSun$
with a $20\%$ uncertainty.
\begin{table}[]
\center
\caption{Equivalent widths (in $\AA$) of the most significant emission lines.}
\begin{tabular}{cccc}
\hline
Line ID & EW$_{1}$ & EW$_{2}$ & EW$_{3}$ \\
\hline
\hline

Ca II K         & 25.90$\pm$3.00   & 21.25$\pm$3.10   & 23.30$\pm$2.50 \\
Ca II H         & 15.70$\pm$3.10   & 12.40$\pm$3.00   & 13.40$\pm$3.00\\
$[$SII$]$      & 0.24$\pm$0.10     &  0.30$\pm$0.10    & 0.21$\pm$0.10\\
H$\beta$     & 1.80$\pm$0.30     & 1.85$\pm$0.30    & 1.53$\pm$0.30\\
He I             & 0.12$\pm$0.06     &  0.14$\pm$0.06    & 0.10$\pm$0.06\\
$[$OI$]$      & 0.12$\pm$0.02     &  0.11$\pm$0.02    & 0.10$\pm$0.02\\
$\Ha$          & 2.96$\pm$0.20     & 3.16$\pm$0.20    & 2.42$\pm$0.20\\

\hline
\end{tabular}
\tablefoot{The subscripts indicate the first (2012-01-09), second (2012-02-24), and third (2012-03-07) nights of the UVES observations (Table~\ref{tab:tab_uves_log}).
Error bars represent the $3\sigma$ uncertainty.}
\label{tab:tab4}
\end{table}

\begin{figure}[]
  \centering \includegraphics[width=\columnwidth, trim = 0 40 0 0]{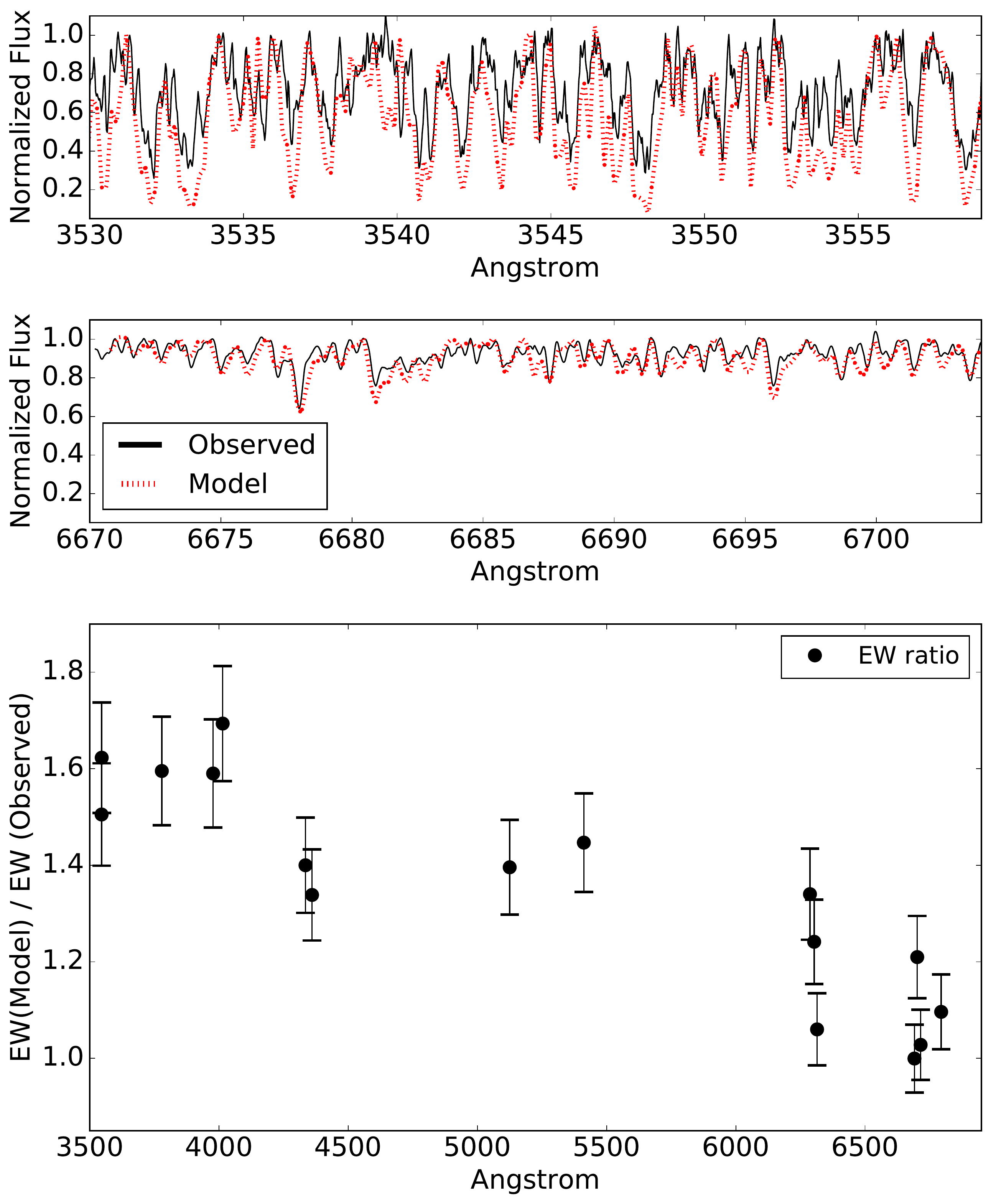}
  \caption{{\bf Top:}  Spectrum of DZ Cha in two different regions (black solid line) compared with the synthetic model (red dashed line). 
  At red wavelengths the spectrum and model match each other and the lines have similar depths in both of them; at blue wavelengths the absorption
  lines of the spectrum are shallower than those in the model. {\bf Bottom:}  Ratio of the equivalent widths of some selected lines across the
  spectrum plotted against the corresponding wavelengths. The ratios model/observations decrease as the wavelength increases. See text for details.}
   \label{fig:fig_veil}
\end{figure}

A comparison between the synthesized stellar photosphere and the UVES spectra shows that the observed absorption lines are shallower than
the model at wavelengths $\lesssim 6500\,\AA$, while model and observations are in excellent agreement at longer wavelengths (Fig.~\ref{fig:fig_veil},
left panel). To further quantify this effect, we measured equivalent widths of some selected lines. In Fig.~\ref{fig:fig_veil} (right panel) the ratio of
the equivalent widths of the model and observations for several absorption lines at different wavelengths are shown. It is clear that the observations
and model are in agreement at wavelengths $\gtrsim 6500 \AA$, while the observations have smaller equivalent widths than the model at bluer
wavelengths. This effect is independent of the broadening factor used to model the synthetic spectra, and at first sight this seems to be the signature of
veiling created by accretion shocks observed in many CTTS \citep[e.g. ][]{1990ApJ...363..654B, 1991ApJ...382..617H}.

\subsubsection{ Ultraviolet excess}
\label{sub_sect:uv_excess}

The SED of DZ Cha (Fig.~\ref{fig:fig_sed}, left panel) shows a prominent UV excess. If this emission is caused
by magnetospheric accretion or magnetic activity, then it can be reproduced by a scaled  black body as
\begin{equation}
F_{UV} = X  \, \frac{R_{\star}^2}{d^2} \, \pi \, B_{\nu}(T_\mathrm{BB}),
\end{equation}
where $B_{\nu}(T)$ is the Planck function at the black-body temperature $T_\mathrm{BB}$, and X is a scaling factor. Fitting
the two GALEX photometric points including distance and extinction uncertainties to this function we find that a black body
at $T_\mathrm{BB} = 12,000_{-2000}^{+3000}$ K can match the UV excess (grey dot-dashed line in Fig.~\ref{fig:fig_sed}, left
panel). With a luminosity of $\lesssim 0.001\,\LSun$
this black body is the dominant source of emission at UV wavelengths, but its contribution to the observed SED quickly becomes
negligible, dropping below $<1\%$ at wavelengths longer than 0.55\mic. 
Assuming that all this flux is powered by accretion, then an independent estimate of the mass accretion rate can be obtained
from the accretion luminosity as $L_\mathrm{acc} \approx 0.8\, G\, M_\star \dot{M}/R_\star$ \citep{1998ApJ...492..323G}. For DZ Cha, this yields to
$\dot{M} < 8.6 \times10^{-11} \MSun \, \mathrm{yr}^{-1}$, where the upper limit reflects that we have ignored the (significant) contribution
from chromospheric activity. Considering all uncertainties and the narrow $\fw \sim 130\,\kms$ discussed above, we adopt an upper limit of
$\dot{M} < 10^{-10} \MSun \, \mathrm{yr}^{-1}$ for DZ Cha.

\subsection{ Circumstellar disc}
\label{sect:disc}

\subsubsection{ Infrared and sub-mm emission}
\label{sect:sed}

\begin{figure*}[]
  \centering \includegraphics[width=\textwidth, trim = 0 30 0 0]{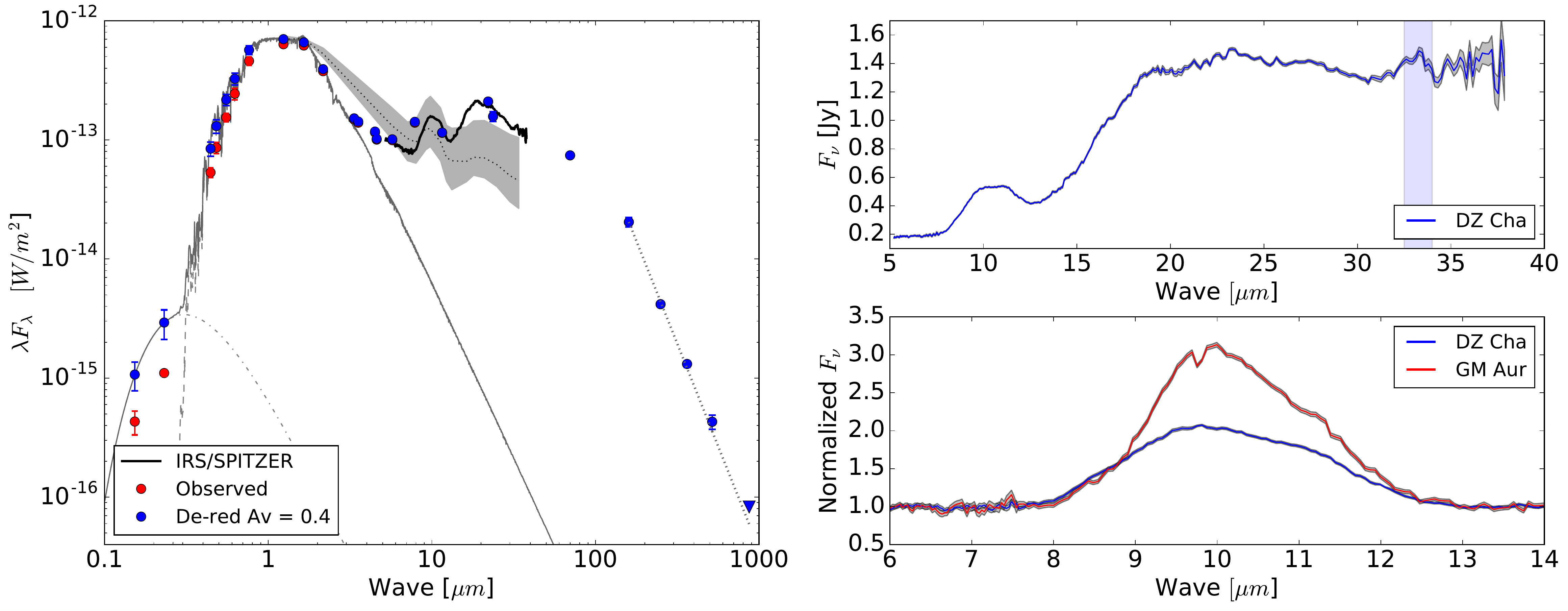}
  \caption{\textbf{Left:} SED. The solid grey line shows the combination of a 12,000 K black body (dot-dashed line) with the photosphere model (dashed line,
  see Sect.~\ref{sub_sect:stellar}). The dotted black line and the grey shadowed region trace the median SED and upper/lower quartiles of Class II discs around
  K2-M5 stars in Taurus \citep{2006ApJS..165..568F}. The dotted grey line shows the power-law fit $F_{\nu} \propto \nu^{\alpha}$, with $\alpha = 2.5\pm0.1$
  (see Sect.~\ref{sect:disc}). \textbf{Top right.} IRS/SPITZER spectra. The light blue rectangle is centred around the 33.8\mic crystalline silicate feature, and
  the grey shadowed region indicates the spectra uncertainties. \textbf{Bottom right.} De-redened and normalized 10\mic profile of DZ Cha and GM Aur.}
   \label{fig:fig_sed}
\end{figure*}

The SED and SPITZER/IRS spectra are plotted in Fig.~\ref{fig:fig_sed} (left panel), along with the J-band normalized photosphere
model described in the previous section. In the near-IR the SED is essentially photospheric at wavelengths shorter than 3.3\mic.
Compared to the median emission of the K2-M5 stars with Class II SEDs (i.e. bright discs with strong thermal emission, most of
them CTTS) observed in Taurus \citep{2006ApJS..165..568F}, DZ Cha shows a clear deficit in flux up to $\sim6$\mic. This is the
characteristic signature of the transitional discs \citep{1989AJ.....97.1451S}. As already noted by these authors, this probably indicates
dust-depleted central cavities of discs in the process of inside-out dispersal. The red IRAC colours
$[3.6] - [4.5] = 0.53$ and $[4.5] - [5.8] = 0.59$ indicate that this cavity is not entirely devoid of dust \citep{2005ApJ...629..881H}.
Beyond 6\mic the SED shows robust thermal emission, with a prominent flux increment reaching the maximum at 
$\lambda \sim 20$\mic (corresponding to $\sim 145$ K). This near- and mid-infrared SED shape is similar to that of GM Aur,
where the moderate NIR excess is explained by a small amount of hot dust and the steep rising in the continuum at IRS wavelengths
indicates an inner wall directly exposed to the stellar radiation at temperature $T_\mathrm{wall} = 130$ K  \citep{2005ApJ...630L.185C}.
Assuming that the dust grains in the wall re-emit the received energy from the star as a grey black body, a crude estimate of the
wall location in DZ Cha can be derived from $d_\mathrm{wall} = (R_{\star}  \, T_{eff})^2  / ( 2 \, \sqrt{\epsilon} \, T_\mathrm{wall}^2)$,
where $\epsilon$ is the dust emissivity ($\epsilon = 1$ for a black body). For dust grains $\epsilon$ is very uncertain and depends
on the grain properties. Using $T_\mathrm{wall} = 145$ K and $\epsilon = [0.1 - 0.9]$ results in $d_\mathrm{wall} = 3-8$ au.
At long wavelengths ($\lambda \ge 150$\mic)
the disc is expected to become more and more optically thin \citep[e.g. ][]{2016A&amp;A...586A.103W} and its emission can be
approximated as a black body radiating in the Rayleigh-Jeans regime, with $F_{\nu} \propto \nu^{\alpha}$. Fitting the observed
$F_{\nu}$ fluxes at these wavelengths to a power law yields $\alpha = 2.5\pm0.1$ (see the dotted line in the left panel of
Fig.~\ref{fig:fig_sed}). This corresponds to a dust opacity index $\beta = \alpha - 2 = 0.5$, much lower than the ISM $\beta \sim 1.7$
\citep{1999ApJ...524..867F, 2001ApJ...554..778L}. This difference indicates emission from large, probably mm-sized and above, dust
grains \citep{2006ApJ...636.1114D, 2014prpl.conf..339T}. Assuming that the continuum emission at $870\,\mu$m is optically thin 
and a representative average dust temperature the total dust mass can be estimated as

\begin{equation}
M_\mathrm{dust} = \frac{F_{\nu} \, d^2}{\kappa_{\nu} \, B_{\nu}(<T_\mathrm{dust}>)},
\end{equation}
\citep{1983QJRAS..24..267H}, where $F_{\nu}$ is our $3\sigma$ upper limit at frequency $\nu = 345$ GHz (870\mic), $d$ is the distance to DZ Cha
($d = 110$ pc), $\kappa_{\nu}$ is the dust opacity, and $B_{\nu}(<T_\mathrm{dust}>)$ is the Planck function for a black body
emitting at the average dust temperature, which is estimated following \cite{2013ApJ...771..129A} as
$<T_\mathrm{dust}> \, = 25\,(L_{\star} / L_{\sun})^{0.25} = 22 $ K. We assume that the dust opacity is a power law in frequency
$\kappa_{\nu} \propto \nu^{\,\beta}$ normalized to 0.1 cm$^2$g$^{-1}$ at 1000 GHz for a gas-to-dust mass ratio of 100:1
\citep{1990AJ.....99..924B}. Using the previously derived opacity index of $\beta = 0.5$ results in $\kappa_{345\mathrm{GHz}} = 2.8$ cm$^2$g$^{-1}$,
which in turn yields a dust mass upper limit $M_\mathrm{dust} < 3\,M_\mathrm{Earth}$. Using a higher opacity index of e.g. $\beta = 1.0$
results in an even lower upper limit ($M_\mathrm{dust} < 2.5\,M_\mathrm{Earth}$), so we are confident about the disc being mostly optically
thin at long wavelengths.

The IRS spectrum, shown in detail in Fig.~\ref{fig:fig_sed} (right upper panel), shows a broad emission feature at 9.7\mic typical of amorphous
grains composed by silicates. Furthermore, at $\sim$ 33.8\mic, there is clearly  a feature at the wavelength associated with crystalline forsterite
\citep{2003A&amp;A...399.1101K}. For comparison, we have normalized the continuum around the 10\mic region for DZ Cha and for the
protoplanetary disc GM Aur (Fig.~\ref{fig:fig_sed}, right bottom panel). In GM Aur, the shape of this feature indicates submicron sized (pristine)
ISM grains \citep{2006ApJ...645..395S}. In DZ Cha, the 10\mic feature points to dust that is more evolved than that of GM Aur: the red shoulder,
typical of larger ($\gtrsim$ 2.0 micron grains) is more pronounced, and the peak-to-continuum ratio is 2.07, while it is 3.13 for GM Aur. This
peak/continuum ratio is shown to decrease as the dust grains grow \citep{2003A&A...400L..21V}. It is thus clear that dust processing has taken
place in the disc of DZ Cha, as also indicated by the slope of the SED in the far-infrared and sub-mm ranges.
There is no evident emission from the [Ne II] 12.8\mic fine structure line. We have fit the continuum around the line to a second-order
polynomial and subtract it to the spectra, finding a tentative ($\lesssim2\sigma$) peak in emission at the line central wavelength. Given its low
significance we give the $3\sigma$ upper detection limit, which yields a line luminosity of $L_\mathrm{[Ne II]} <
6.4\times10^{28}$ ergs s$^{-1}$ ($<1.7\times10^{-5}\,\LSun$).

\subsubsection{The disc in polarized light}
\label{sect:pol_images}

\begin{figure*}[]
  \centering \includegraphics[width=\textwidth, trim = 0 80 30 0]{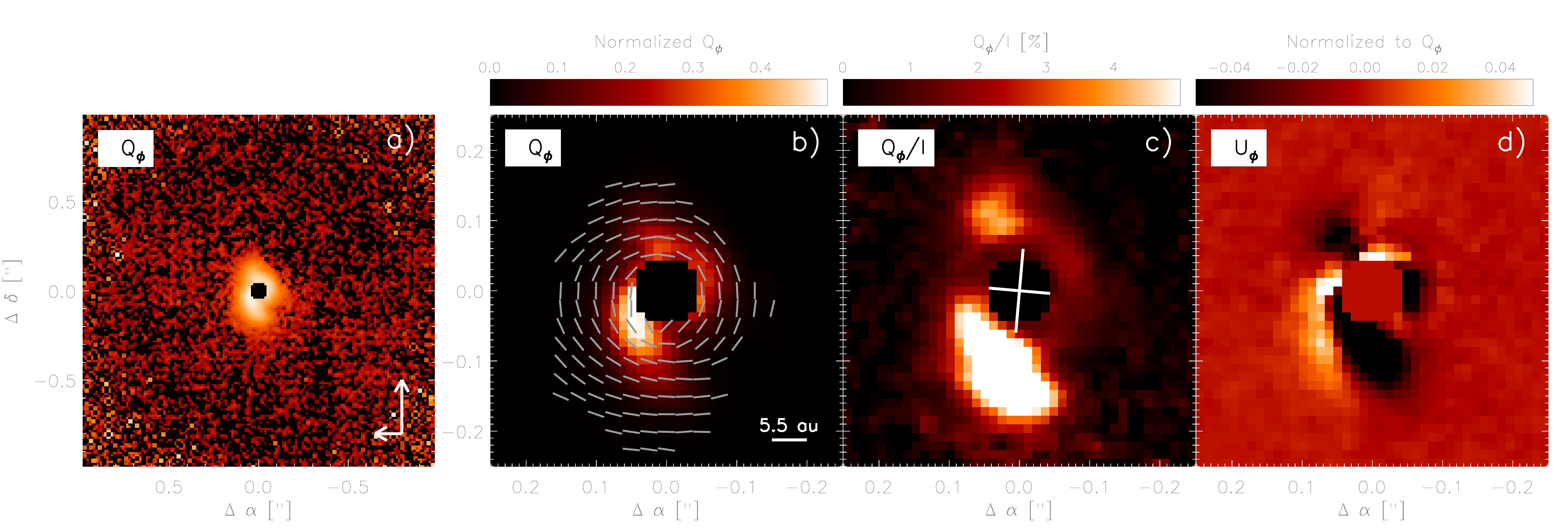}
  \caption{\textbf{a)}. $Q_{\phi}$ image of DZ Cha in  logarithmic scale. The rest of the panels are shown in  linear scale and the dynamical range
  of each image has been adjusted to better show the disc structure. \textbf{b)}. Zoom-in of a)
  with the polarization angle $P_{\theta}$ plotted as grey vectors. \textbf{c)} Ratio of $Q_{\phi}$ over the total intensity ($I$). The central cross
  indicates the projected major and minor disc axis. \textbf{d)} $U_{\phi}$ image normalized to the maximum value of $Q_{\phi}$. The positive/negative
  pattern is  characteristic of multiple scattering events. In all panels the innermost region ($r < 0\farcs04$) dominated by speckle noise
  has  been masked out;  north is up and east is left.}
   \label{fig:fig_sphere}
\end{figure*}
\begin{figure}[]
  \centering \includegraphics[width=0.8\columnwidth, trim = 0 20 0 15]{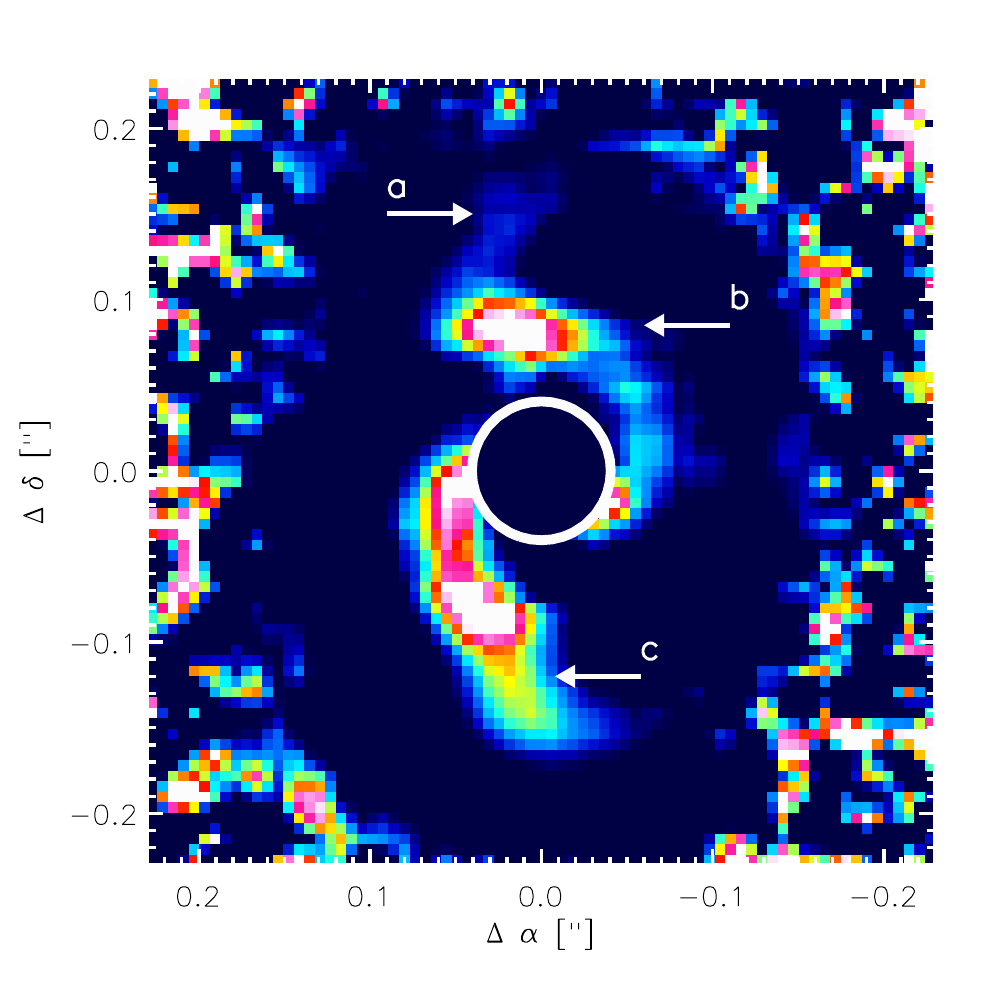}
  \caption{Unsharped $Q_{\phi}$ image plotted in a hard stretch and colour scale. The inner $r < 0\farcs04$ region (white circle) is masked out. The arrows
  indicate different features discussed in the text.}
   \label{fig:fig_uns_mask}
\end{figure}

Our images, which are sensitive to the small (\mic-sized) dust grains that populate the disc upper layer, reveal a complex circumstellar
environment. Polarized emission is detected in both the coronagraphic and non-coronagraphic observations. The coronagraphic images
are blurrier than the unmasked ones, probably because of the different weather conditions that translated into a better AO performance in
the unmasked dataset. The average point spread function (PSF) of the unmasked observations is almost diffraction limited with a FWHM of $\lesssim 0\farcs05$,
and in this dataset polarized emission features are detected well inside the region masked out by the focal mask.
Therefore, we focus our analysis on the unmasked observations (Fig.~\ref{fig:fig_sphere}). The polarization angle $P_{\theta}$ is distributed
in a nearly azimuthally symmetric pattern consistent with scattering by dust on a disc surface. The images show two elongated structures
reminiscent of spiral arms in an m = 2 rotational symmetry, with the northern (fainter) one extending from azimuthal angles [$270\degr-20\degr$] (east of north), and
the southern (brighter) one covering azimuthal angles [$100\degr-190\degr$] (see central panels in Fig.~\ref{fig:fig_sphere}). The $U_{\phi}$ image
(panel d) shows a positive and negative pattern with a maximum value of $\sim 20\%$ of the $Q_{\phi}$ image (peak to peak), and on average
its amplitude is within $\sim 10\%$ of $Q_{\phi}$. This is consistent
with multiple scattering events in an inclined ($i \ge 40\degr$) disc \citep{2015A&amp;A...582L...7C}. To highlight the disc morphology
we applied different unsharp masks to the $Q_{\phi}$ images as in \cite{2016A&A...588A...8G}. The spiral-like features appear in every filter applied,
showing in all cases a very similar morphology (Fig.~\ref{fig:fig_uns_mask}). We also highlight two features (labelled  `a' and `c') that seem spatially
connected to the outer end of the spirals , and a bright clump (feature `b') in the northern spiral. Hints of these features are observed in the $Q_{\phi}/I$
panel in Fig.~\ref{fig:fig_sphere}.

A first estimate of the disc structure can be obtained assuming that the disc is centro-symmetric and geometrically flat. We begin by smoothing the
$Q_{\phi}$ image with a 2 px  Gaussian kernel to reduce the impact of the bright and isolated
features. We then fit isophotal ellipses to three  different surface brightness ranges excluding the central ($r < 0\farcs04$) and outer ($r > 0\farcs15$)
disc regions that are dominated by noise. This procedure was repeated using Gaussian kernels of increasing width (up to 10 px). This way we derive
a position angle of PA = $176\degr \pm7\degr$ and an inclination of $i = 43\degr \pm 5\degr$, where the quoted errors correspond to the standard
deviation ($1\sigma$) of the different fits. We note that caution must be taken with our result as the disc is certainly not symmetric and probably flared
(see Sect.~\ref{sub:a_complex}).

\begin{figure}[]
  \centering \includegraphics[width=1.0\columnwidth, trim = 20 30 20 0]{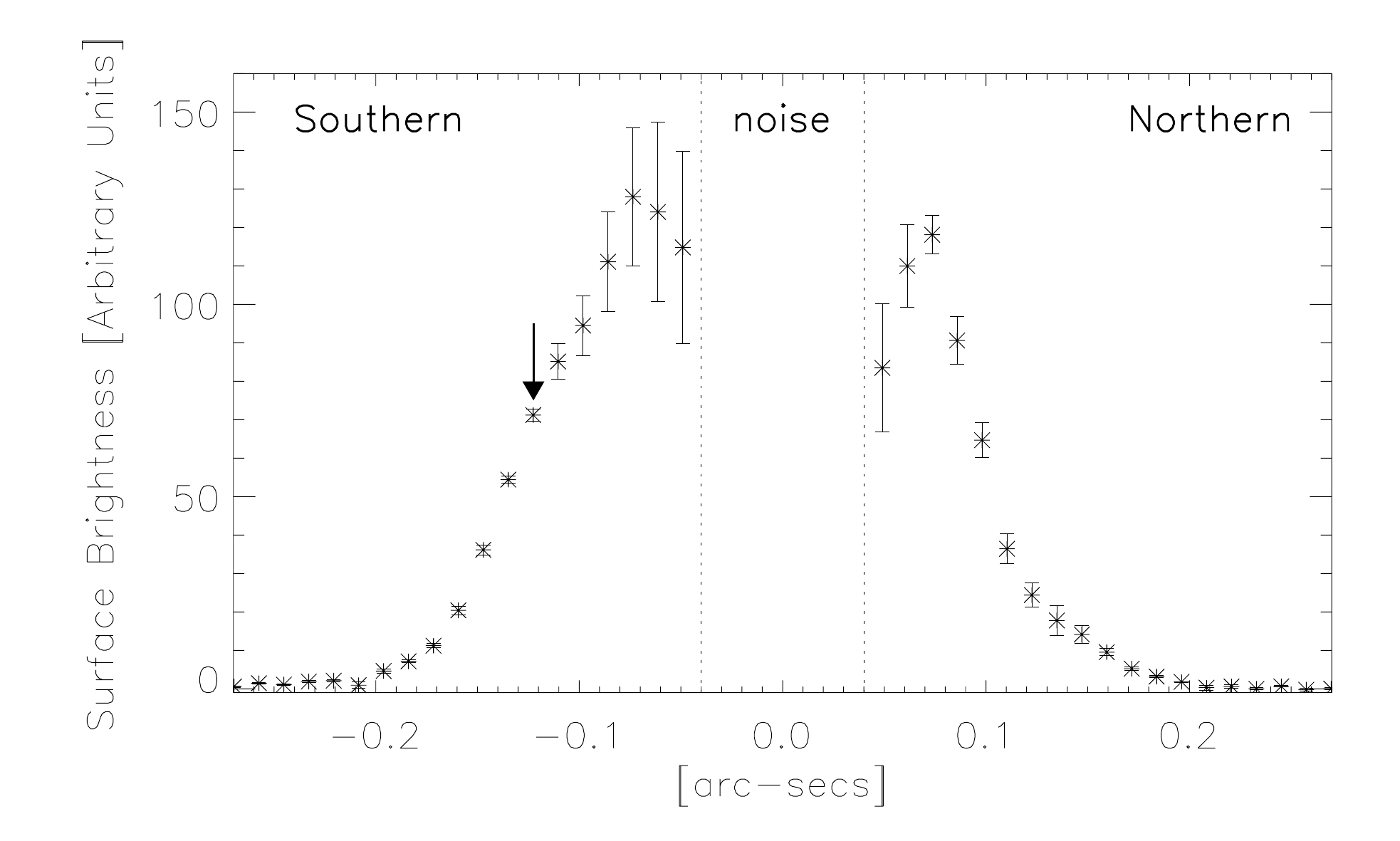}
  \caption{Radial profile along the disc major axis. The southern and northern side peak at $\sim 0\farcs07\pm0\farcs01$. The arrow shows the radius where the
  brightness profile changes in slope along the southern side.}
   \label{fig:fig_rad_prof}
\end{figure}

The radial (polarized) brightness profile was measured by computing the mean in a 3 px slit ($\sim 1$ resolution element)  along the disc
major axis (Fig.~\ref{fig:fig_rad_prof}). The uncertainties at each position are defined as the standard deviation divided by the square root
of the slit width. The two semi-major axes peak at $0\farcs07 \pm 0\farcs01$ and show an abrupt decrement in emission towards the star.
This is the typical signature of a disc inner cavity.
The semi-major axes are not symmetric, and overall the southern axis is brighter than the northern counterpart. The southern side is detected above $3\sigma$
up to $0\farcs23\pm0\farcs01$ ($\sim25$ au) from the star, while the northern side is detected up to $0\farcs18\pm0\farcs01$ ($\sim20$ au).
Neither of the two sides can be fitted by a single power law $\propto r^{\alpha}$, and the southern side shows a pronounced change in brightness
at $0\farcs12\pm0\farcs01$ ($\sim 13$ au).

The dilution effect of the PSF can create a fake small central hole in polarized light \citep{2014ApJ...790...56A}. This artificial cavity depends
on the PSF properties and, for example, a fake inner hole at $0\farcs05$ can be observed in diffraction limited observations at Ks band.
To test whether the observed emission decrement near the star is an artefact we ran a set of simple toy models using the radiative transfer code MCFOST
\citep{2009A&A...498..967P}. We created a continuous disc model (down to 0.1 au from the star)   and a set of discs sharply truncated at an inner
radius ranging from 5 to 9 au. We used a central star with the same properties as DZ Cha, a standard power law $\Sigma(r) \, \propto r ^{-1}$ to
describe the disc surface density distribution, a disc inclination of $43\degr$,
and a distance of 110 pc. Other disc properties such as grain composition or scale height have no impact for the purposes of this test. The modelled
Stokes Q and U images were convolved with the average observed PSF, and synthetic $Q_{\phi}$ images were obtained from those. We then
computed the cuts along the projected major axis as we did with the observations. Comparing the synthetic cuts with the observed one we found
that the decrement in polarized flux is not an artefact but a real inner cavity of $7\pm0.5$ au in radius. This cavity size is in agreement with the inner
wall location derived from our analysis of the SED shape (see Sect.~\ref{sect:sed}). 

\subsubsection{Companion detection limits}
\label{sect:det_limits}

\begin{figure}[]
  \centering \includegraphics[width=\columnwidth, trim = 0 30 0 0]{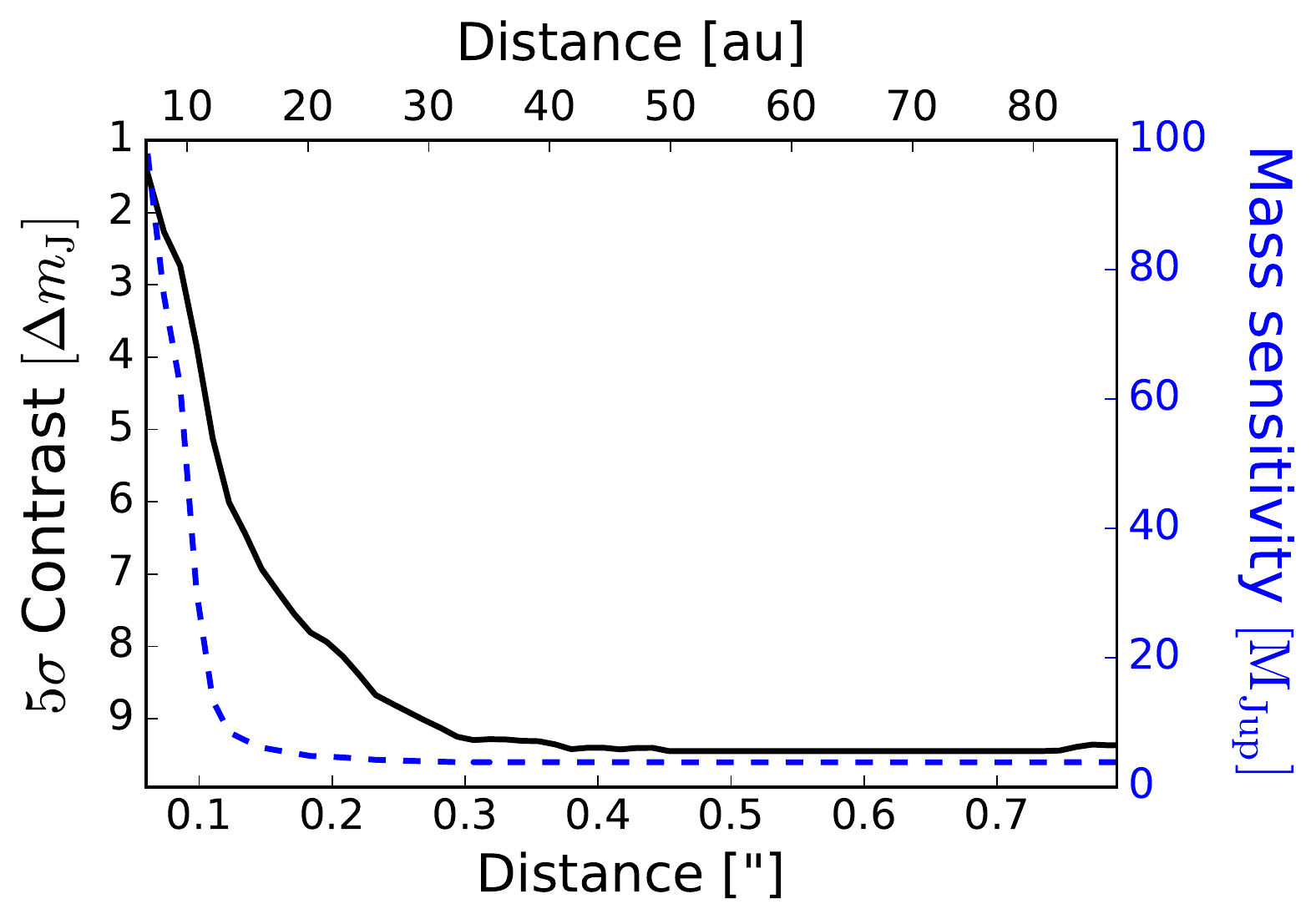}  
  \caption{\textbf{Limits derived from the $J$-band intensity images}. The solid black line traces the $5\sigma$ contrast limit with its scale given on the y-left axis.
  The dashed blue line shows the mass sensitivity limits derived using AMES/Dusty models \citep{2001ApJ...556..357A},
  with its corresponding scale on the y-right axis.}
   \label{fig:fig_contrast}
\end{figure}

The average $5\sigma$ detection limits derived from our intensity images are shown in the left panel of Fig.~\ref{fig:fig_contrast}. Beyond $0\farcs2$ (22 au
projected) the contrast drops below 8 magnitudes, reaching a roughly constant value of $\Delta \, m_{J} \sim 9.5$ magnitudes at separations larger
than $0\farcs4$. Using the distance and J-band magnitude of DZ Cha ($m_\mathrm{J} = 9.5$), our detection limits imply that we should detect objects
with absolute magnitude at J band $M_\mathrm{J} \lesssim 12.3$ at separations larger than $0\farcs2$. We used the AMES/Dusty models for brown
dwarfs and giant planets atmospheres \citep{2001ApJ...556..357A} to derive mass sensitivity limits from our detection limits. The predicted temperature
(and therefore brightness) depends on the assumed age, so for this exercise we use an age of 2.2 Myr (i.e. as if the companion were coeval with the
central star). The mass sensitivity limits derived in this way are shown in the right panel of Fig.~\ref{fig:fig_contrast}. Giant planets with masses above
$5M_\mathrm{Jup}$ orbiting beyond 22 au (projected) could be detected in our observations, although we note that  the uncertainties are probably
very high given the sensitivity of the planet brightness to the initial conditions of planet formation \citep{2013A&amp;A...558A.113M} and that the disc
emission may mask the planet signal. Our observations rule out stellar companions ($M_\star > 80 \MJup$) down to $0\farcs07$ (projected $\sim 8$ au)
and equal mass companions down to $0\farcs05$ (projected $\sim 5$ au).

Close-in stellar companions can be detected by combining several epochs of high resolution spectra as significant radial velocity variations will appear
depending on the mass ratio and separation of the binary. \cite{2014A&amp;A...568A..26E} used this method to identify spectroscopic companions in a
large sample including DZ Cha. They do not find evidence for binarity in DZ Cha using the same UVES dataset discussed here. Their result is consistent,
within error bars, with previous measurements covering a time baseline of $\sim20$ years \citep{1997A&amp;A...328..187C, 1992AJ....103..549G, 2006A&amp;A...460..695T}.
To further test whether the various radial velocity measurements of DZ Cha show statistically significant radial velocity variations we performed a $\chi^2$ test.
As a null hypothesis we assumed that the average of the radial velocity measurements is a good representation of the constant radial velocity of DZ Cha.
Using the $\chi^2$ probability function we find that the available observations do not provide any evidence for radial velocity variations. Combining the
six independent measurements found in the literature results in an average radial velocity of $13.9 \pm 0.5\, \kms$ ($1\sigma$ error computed including
the individual uncertainties). In a binary system the radial velocity semi-amplitude ($K = v_{r, max} - v_{r, min}$) and the semi-major axis ($a$) of the orbit
are related via the Kepler laws. If the orbit is circular then $K = \sqrt{G} \, m_2 \, \sin i \, (m_1 + m_2)^{-0.5} \,a^{-0.5}$. Assuming that the orbit is coplanar
with the disc ($i = 43\degr$), and using a primary mass of $m_1 = 0.7\MSun$ and a $3\sigma$ detection limit of $1.5\,\kms$, the current measurements
suggest that companions above $0.2\MSun$ at all separations are unlikely.

\section{Discussion}
\label{sec:discusion}

\subsection{Accretion or chromospheric emission?}
\label{sec:discusion_1}

Many of the observed emission lines require temperatures well above the temperature of the stellar photosphere. In particular,
the He I 5876$\,\AA$ line only shows up in emission when the temperature exceeds 8,000 K \citep[e.g. ][and references therein]
{1997A&A...326..741S, 2001ApJ...551.1037B}. Furthermore, the spectra also show veiling signatures (Sect.~\ref{sub_sect:em_lines}).
In TTSs two mechanisms can produce the high energies needed to create these features: accretion and magnetic activity.

In young stars, the Balmer lines, the He I, and the Ca II lines can originate from a very hot inner wind or from an accretion shock
\citep{1996ApJS..103..211B, 1998ApJ...492..743M, 2001ApJ...551.1037B, 2002A&A...391..595S, 2011MNRAS.411.2383K},
while veiling is usually considered  a signature of accretion activity, \citep[e.g. ][]{1998ApJ...492..323G, 1998ApJ...495..385H}. However,
the narrow $\Ha$ line profile argues against accretion in DZ Cha. 
The $\Ha$ line is known to be variable in TT stars \citep[with the $\fw$
having a typical dispersion of 0.65 dex;][]{2009ApJ...694L.153N}, but our three different spectra and the observations from \cite{2010ApJ...724..835W}
show a nearly constant $\fw < 130\,\kms$. Similarly, combining our three spectra with observations at five different epochs
from the literature we find that the equivalent width  consistently remains  around $\ew = 4\pm2\,\AA$ in timescales from months
to years, except during two flaring events \citep{1992AJ....103..549G, 1995A&amp;AS..114..109A, 1997A&amp;A...321..803G, 
2000MmSAI..71.1021T, 2006AJ....132..866R, 2006A&amp;A...460..695T}. According to \citet{1998MNRAS.300..733M}, M0-M2
stars with $\ew$ > $10\AA$ are accreting. \citet{2003ApJ...582.1109W} find that CTTS have $\mathrm{FW0.1}(\Ha) > 270\,\kms$,
independently of their spectral type, and this limit drops to $200\,\kms$ for substellar objects \citep{2003ApJ...592..282J}. \citet{2003AJ....126.2997B}
propose an empirical criterion based on the saturation limit traced by $\log  [L(\Ha) / L(\mathrm{bol})] = -3.3$. In this scenario,
M0 CTTS have $\ew > 8\AA$ (with small variations depending on the observed star forming region, see their Fig. 4). In short the $\Ha$
line profile indicates no accretion in DZ Cha, as this object is not even in the boundary between WTTS and CTTS in any of these widely
used empirical criteria. Furthermore, the He I 5876$\,\AA$ line profile also argues against accretion in DZ Cha. Accreting objects usually
have larger equivalent widths in the He I line (EW(He I)) than non-accreting ones. The EW(He I) in DZ Cha is $\sim30\, \times$ lower than
the EW(He I) of the accreting K7-M2 PMS stars studied by \cite{2014A&amp;A...561A...2A}, and is $\sim3\, \times$ lower than
the EW(He I) of the non-accreting K7-M2 PMS stars in the sample studied by \cite{2013A&amp;A...551A.107M}.

An alternative and plausible explanation for the UV excess, veiling, and emission lines is chromospheric activity. Ultraviolet excesses caused by
chromospheric emission alone have  already been observed in a number of WTTS and magnetically active M dwarf stars
\citep{1996A&amp;A...305..209H, 2013ApJ...767..112I}. DZ Cha is magnetically active as it shows flaring events \citep{1997A&amp;A...321..803G,
2000MmSAI..71.1021T}. Therefore, the UV excess observed in DZ Cha could be caused by just chromospheric emission,
and this excess may account for the observed veiling signatures. Similarly, high transitions of the Balmer series and several emission lines
observed in DZ Cha (e.g. He I, $\Ha$, Ca I) are also observed in WTTS due to chromospheric emission alone \citep{2013A&amp;A...551A.107M}.
Finally, it could be argued that short episodes of enhanced accretion cause outbursts that mimic the observational signatures
of the stellar flares. However, young outbursting objects such as FU Ori and EX Ori stars have average accretion rates in quiescence of
$\sim 10^{-7}\,\MSun \, \mathrm{yr}^{-1}$, and their outbursts increase the stellar flux at optical wavelengths by a few magnitudes
over at least several months \citep[][and references therein]{2014prpl.conf..387A}. This is clearly not the case in DZ Cha, so the two
flaring events observed to date in this system indeed indicate intense magnetic activity. 

Therefore, and taking into account the previous discussion about
the $\Ha$ line and accretion variability, we conclude that magnetic activity, and not accretion, is the mechanism heating the stellar
chromosphere and causing the observed UV excess, emission lines, and veiling.

\subsection{A complex and evolved disc}
\label{sub:a_complex}

Our observations show that DZ Cha has undergone significant evolution from its primordial stage, where dust grains in the circumstellar
disc should be ISM-like and the inner disc should extend down to the dust sublimation limit \citep[<0.5 au for TTS, ][]{2003ApJ...597L.149M}.
In DZ Cha, the IRS spectrum presents the signature of crystal silicates at $\sim 33.8$ \mic, while the shape of the 10\mic feature indicates
$\gtrsim$ 2.0\mic dust grains in the optically thin upper layers of the disc. The spectral index $\beta \sim 0.5$ derived from the far-infrared and sub-mm
SED probes the continuum emission from large, likely mm-sized, grains. Furthermore, the strong infrared excess indicates that the disc is
optically thick at most infrared wavelengths. Our polarized images and in particular the brightness surface profile along the disc major axis
show a dust cavity (Fig.~\ref{fig:fig_rad_prof}). Taking into account the dilution effect of the PSF we find that this cavity is $\sim7$ au in radius. 
An optically thin dust-depleted cavity is also evident from the transitional
disc-like SED, which shows no excess below 3.3\mic and has a prominent increment from 6 to 20\mic characteristic of an inner wall directly
exposed to the stellar radiation. The relatively small outer radius of $\sim 30$ au from our observations suggests a compact disc, but deep
observations at longer wavelengths are needed to accurately constrain the disc size. Our upper mass limit from the $870$\mic photometry
($<3\,M_\mathrm{Earth}$, Sec.~\ref{sect:sed}) indicates a total mass $M_\mathrm{disc} < 0.95 \MJup$ assuming a conservative gas-to-dust
mass ratio of 100:1. Using our derived stellar mass ($M_\star = 0.5\MSun$, Table~\ref{tab:tab_stellar_props}), we obtain an upper limit for the
star-disc dust mass ratio of $M_\mathrm{disc}/M_\star < 0.2\%$. 

The spiral features observed in the polarized images seem to be in an approximate m = 2 rotational symmetry, hinting at an undetected
companion. The theory shows that if this is the case, the observed morphology depends on several factors including disc properties (like
the $\alpha$ viscosity parameter, scale height, and inclination), the star/companion ratio $q$, and the companion location
\citep[e.g.][]{2015ApJ...809L...5D, 2015MNRAS.451.1147J}. Current models predict that when observed in NIR scattered light images,
\textit{outward} spirals (those launched beyond the companion) are more difficult to detect than \textit{inward} spirals \citep{2015ApJ...809L...5D,
2015ApJ...813...88Z, 2017ApJ...835...38D}. Independently of the companion mass, outward spirals are expected to have very small pitch angles,
in contrast to the open arms predicted when the spiral is located between the companion and the star. Therefore, the bright and open spirals
observed in DZ Cha suggest inward  spirals launched by a companion orbiting in the outer disc. In our images the spirals are not detected beyond
$\sim 20$ au, so we adopt this value as the minimum orbit of the hypothetic companion. The minimum mass of a companion driving a detectable
spiral feature in NIR scattered light depends on the detailed properties of the disc, and for DZ Cha it might be around
$\sim 0.1 \MJup$ \citep{2017ApJ...835...38D}. The morphology and brightness of the $Q_{\phi}$ and $U_{\phi}$ images is roughly consistent with
an inclined generic disc with inward spiral arms driven by a $0.5 - 50\MJup$ companion ($q = 0.003, 0.1$) \citep[][]{2016ApJ...826...75D}. Our
point-source detection limits suggest that such a companion should be $\lesssim 5 \MJup$. Altogether we conclude that a companion with
masses in the $0.1-5\,\MJup$ range driving inward spirals and orbiting at $r_\mathrm{orbit} >20$ au might explain the observed morphology of
DZ Cha.

The lack of accretion does not necessarily imply a gas-poor disc and, in fact, there is indirect and direct observational evidence of gas in DZ Cha.
The infrared excess of DZ Cha is much stronger than the excesses found around gas-poor discs \citep[e.g. ][]{2013A&amp;A...555A..11E}. The
overall SED shape and disc fractional luminosity of DZ Cha are, however, very similar to those observed in the flared gas-rich discs around CTTS.
Direct evidence of the presence of gas in the disc comes from our detection of the forbidden [OI] and [SII] emission lines in all  three epochs. These
lines are usually detected in CTTS, and their low velocity in our spectra indicate that they trace emission from warm gas in the disc surface
\citep[e.g. ][]{1995ApJ...452..736H, 2014A&A...569A...5N}.

\subsection{A bona fide photoevaporating disc}

Mechanisms like grain growth, viscous accretion, photoevaporation, and planet formation, are expected to play different
roles in the disc evolution depending on the intrinsic properties of the star$+$disc system \citep[see e.g.][]{2014prpl.conf..497E}. This is probably reflected
on the diversity of SEDs observed in the WTTS population \citep[e.g.][]{2013ApJ...762..100C}. Below we discuss how the properties of DZ Cha argue against
most disc evolution mechanisms except photoevaporation. We then show that the current observational evidence indicates that DZ Cha is most likely a bona fide
photoevaporating disc at the inner stages of disc clearing.

In Sect.~\ref{sect:disc} we showed that grain growth is happening in DZ Cha. Grain growth alone can explain small (<10 au) disc inner cavities
\citep{2012A&amp;A...544A..79B}, but this process cannot explain the negligible accretion rate observed in DZ Cha across different epochs (see
Sect.\ref{sec:discusion_1}). Giant planets ($\ge 1 \MJup$) carve gaps and/or cavities in the dust and gas distribution of protoplanetary discs
\citep{1999ApJ...526.1001L, 2012A&amp;A...545A..81P}. These planets can decrease the accretion rate onto the star by  $10\% - 20\%$
 \citep{2006ApJ...641..526L}. Considering the typical accretion rates of CTTS \citep[$10^{-7} - 10^{-9}\, M_{\sun}$ yr$^{-1}$, ][]{1998ApJ...492..323G},
the low upper limit for the accretion rate in DZ Cha cannot be accounted for by one giant planet alone. Multiple and vigorously accreting giant planets may
reduce the stellar accretion rate to $\lesssim 10^{-10}\, M_{\sun}$ yr$^{-1}$, but then very large disc cavities should be created \citep{2011ApJ...729...47Z}.
Therefore, planet--disc interactions cannot explain the observed properties of DZ Cha. Poorly ionized disc regions (dead-zones) can \textit{locally}
have low accretion rates creating gaps in the disc, but the stellar accretion rate is expected to remain insensitive to these dead-zones
\citep{1996ApJ...457..355G}, and therefore this mechanism cannot explain negligible accretion rates. Finally, binary stellar companions can
carve inner cavities of $\sim2-3\,a$ in radius in their circumbinary discs, where $a$ is the semi-major axis of the binary orbit \citep{1994ApJ...421..651A}.
The impact of binarity in the accretion rate is not yet well quantified, but there are both theoretical and observational studies showing that
the accretion flow continues via accretion streams through the dust cleared central cavity of circumbinary discs \citep{1996ApJ...467L..77A, 1997AJ....114..781B}.
Most importantly, none of the mechanisms enumerated above can explain the disc outflow evidenced by the detection of the forbidden
[SII] and [OI] lines in the UVES spectra.

The shape and luminosities of the forbidden lines, and the stellar X-ray luminosity, can be used to directly compare model predictions
with observations. In Sect. \ref{sect:sed} we obtained $L_\mathrm{[Ne II]} < 6.4\times10^{28}$ ergs s$^{-1}$ ($<1.7\times10^{-5}\,\LSun$).
Taking into account previous observations \citep[e.g. ][]{2007ApJ...665..492L, 2009ApJ...702..724P}, it seems plausible that the IRS
spectra  discussed here just do not have enough sensitivity to detect this line. To further explore this avenue we compared our observations
with the X-ray and EUV photoevaporation models presented by \cite{2010MNRAS.406.1553E}. Scaling the DZ Cha X-ray luminosity given
by \cite{2014ApJ...788...81M} to 110 pc we obtain $\log(L_X) = 30.09$ erg s$^{-1}$. \cite{2010MNRAS.406.1553E} compute several line
luminosities for stars with X-ray luminosities in the $\log L_X = 28.3 - 30.3$ erg s$^{-1}$ range. From these models we estimate a [Ne II]
luminosity of $L_\mathrm{[Ne II]} \lesssim 3.2\times10^{-6}\,\LSun$, i.e. about 5 times below the sensitivity limit of our IRS observations. 
The detection limit of $L_\mathrm{[Ne II]} <1.7\times10^{-5}\,\LSun$ is unfortunately too high to derive a meaningful estimate of the
photoevaporative mass loss rate from this line \citep[following e.g.][]{2009ApJ...703.1203H, 2010MNRAS.406.1553E}.
Repeating the same exercise with the [O I] line ($L_\mathrm{[O I]} \sim 3.1\times10^{-6} \LSun$, Sect.\ref{sub_sect:em_lines})
we find a good agreement between observations and models (with expected luminosities in the $L_\mathrm{[O I]} = 1.9\times10^{-6} - 1.2\times10^{-5}$
range for stellar X rays luminosities of $\log L_X = 29.3 - 30.3$). The line peak velocity is also in agreement with those models, which predict
$v_\mathrm{peak} = [-0.75,-14]\,\kms$ for discs with $\sim8$ au cleared inner cavities inclined by $40\degr-50\degr$. The observed
FWHM of $\sim 46\pm4\,\kms$ is, however, $\sim 60\%$ higher than the models' predictions. Mass loss rates in a photoevaporative wind
depend on the energy of the stellar radiation. Using Eq. 9 in \cite{2012MNRAS.422.1880O} we obtain a mass loss rate due to X-ray
photoevaporation of $\dot M_\mathrm{wind} \sim 9.8\times10^{-9}\,\MSun$yr$^{-1}$, which is nearly 2 orders of magnitude higher than
the mass accretion rate derived for DZ Cha (Sect.~\ref{sub_sect:uv_excess}). 

Photoevaporation alone gives an elegant explanation to most of the observed features in DZ Cha. As explained in Sect.~\ref{sec:intro}, the stellar photons
with energies in the X-ray--UV range can heat the gas in the disc surface by injecting the gas molecules with energy enough to overcome the disc gravitational
potential \citep{2001MNRAS.328..485C, 2006MNRAS.369..229A, 2009ApJ...690.1539G, 2009ApJ...699.1639E, 2012MNRAS.422.1880O}.
The inner disc detaches from the outer disc at a critical radius once the mass loss rates due to photoevaporation equal or exceed the mass accretion rates.
For a  $0.5\MSun$ star like DZ Cha, the critical radius from EUV radiation is expected to be at $R_\mathrm{C} \sim 1$ au \citep[][and references therein]
{2014prpl.conf..475A}. When this happens the inner disc quickly drains in its local viscous timescale, the accretion onto the star stops, and the inner wall
of the outer disc is directly irradiated by the central star. This inner wall then shifts outwards and the entire disc dissipates from the inside out in $\sim 10^{5}$
yr \citep{2006MNRAS.369..229A, 2014prpl.conf..475A}. The multi-epoch observations  discussed here indicate a very low accretion rate
($\dot{M} < 10^{-10} \MSun \, \mathrm{yr}^{-1}$) on timescales from months to years. The SED and images indicate a cavity size of $\sim 7$ au in
radius filled with a small amount of optically thin hot dust. This is consistent with the picture previously outlined:  in DZ Cha the inner disc has probably
already drained (explaining the negligible accretion rate) and the inner wall has moved outwards by up to $\sim7$ au. The broad and slightly blue-shifted
[OI] $6300\,\AA$ emission is consistent with a warm flow of photoevaporated gas escaping from the disc surface \citep{2010MNRAS.406.1553E,
2016MNRAS.460.3472E}. Finally, the low disc mass ($M_\mathrm{disc} < 0.9 \MJup$) is consistent with UV-dominated photoevaporation, as these
models predict that the cavity will open when most of the disc mass has already been accreted by the star \citep{2006MNRAS.369..229A}, as opposed
to  the expected higher mass of discs with cavities carved out by giant planets or X-ray photoevaporating discs \citep{2009ApJ...699.1639E,
2012MNRAS.422.1880O}. 

\section{Summary and conclusions}
\label{sec:conclusions}

In this work we analyse three epochs of high resolution optical spectroscopy with UVES, high contrast J-band imaging polarimetry with SPHERE, mid-infrared
spectroscopy with IRS/Spitzer, and the overall SED from the UV up to the sub-mm regime of DZ Cha. Comparing our results with previous observations we obtain
nine epochs of high resolution spectroscopy encompassing time baselines from years to months. We find that $\ew$ and $\fw$ of DZ Cha remain well below the
WTTS/CTTS boundary, indicating that accretion has stopped or is below $\dot{M} < 10^{-10} \MSun \, \mathrm{yr}^{-1}$. Exceptions to this are two reported flaring
events in which the $\Ha$ line was significantly
broadened \citep{1997A&amp;A...328..187C, 1997A&amp;A...321..803G, 2000MmSAI..71.1021T}. The UVES spectra  analysed here show emission lines indicative
of high temperatures (e.g. He I $5876\,\AA$) and veiling-like effects. We show that given the lack of accretion in this system, and its important flaring events, an
active chromosphere is most likely causing the observed veiling, emission lines, and UV excess. The analysis of the SED and IRS spectra reveals that the disc around
DZ Cha shows evidence of grain growth, probably up to  mm sizes and above, and signatures from crystal silicates. Using our APEX photometry we derive an upper
dust mass limit of $M_\mathrm{dust} < 3 M_\mathrm{Earth}$. Our polarized images reveal a $\sim43\degr$ inclined disc with two spiral-like features and a dust
depleted cavity of $\sim7$ au, which is broadly consistent with the near- and mid-infrared shape of the SED. The spiral shape and polarized $Q_{\phi}$ and $U_{\phi}$
images match the model predictions of an inclined disc with spiral arms driven by planetary mass body exterior to the spirals.

Although grain growth is needed to explain the far-infrared SED slope, and planet--disc interactions can explain the observed spiral features, we argue
that the overall observational characteristics of DZ Cha strongly suggest photoevaporation as the mechanism leading the evolution of DZ CHa. Most
importantly, we show that its observed properties are consistent with a disc at the earlier stages of photoevaporation. At these stages the inner disc has
already drained, accretion has stopped, and the inner wall of the outer disc is exposed to stellar radiation \citep[e.g. ][]{2006MNRAS.369..216A}. DZ Cha
is by no means the only bright disc showing a small accretion rate \citep[e.g. ][]{2006ApJ...638..897S, 2006AJ....132.2135S, 2008ApJ...686L.115C}, and
there are even objects that appear to switch between the CTTS and WTTS categories \citep{2004MNRAS.347..937L}. For example, IM Lup and Sz 68 have
low accretion rates, nearly at the WTTS boundary, and SEDs similar to that of a CTTS with \textit{continuous} and \textit{massive} discs \citep{2008A&A...489..633P,
2013ApJ...762..100C}. However, the combination of negligible accretion rates over the last 20 years, the relatively small disc cavity, the forbidden emission lines,
and the low disc mass, all in agreement with photoevaporation models, seem to be a unique feature of DZ CHa. We consider that DZ Cha is an ideal target to
benchmark photoevaporation theories against direct observations.

\begin{acknowledgements}
        The authors thank the referee, Dr. R. Alexander, for his useful comments and suggestions which helped to improve this manuscript.
        H.C. acknowledges useful discussions with J. Caballero and E. Covino. 
        H.C., C.E., G.M., B.M., I.R., and E.V. are supported by Spanish grant AYA 2014-55840-P.
        G.M. acknowledges support from Spanish grant RyC-2011-07920. 
        M.R.S., L.C., C.C., and A.H. acknowledge founding by the Millennium Science Initiative, Chilean Ministry of Economy, Nucleus RC130007. 
        L.A.C. acknowledges support from CONICYT FONDECYT grant  1171246.
        CC acknowledges support from project CONICYT PAI/Concurso Nacional Insercion en la Academia, convocatoria 2015, Folio 79150049.
        This publication makes use of data products from 
        1) the AAVSO Photometric All Sky Survey (APASS), funded by the Robert Martin Ayers Sciences Fund and the National Science Foundation;
        2) the Two Micron All  Sky Survey, which is a joint project of the University of Massachusetts
        and the Infrared Processing and Analysis Center/California Institute of Technology, funded by the National Aeronautics and Space Administration
        and the National Science Foundation; and
        3) the Wide-field Infrared Survey Explorer, which is a joint project of the University of California, Los Angeles,
        and the Jet Propulsion Laboratory/California Institute of Technology, funded by the National Aeronautics and Space Administration.
        
        This research has made use of 1) the services of the ESO Science Archive Facility; 2) the Spanish Virtual Observatory (http://svo.cab.inta-csic.es)
        supported by the Spanish MICINN / MINECO through grants AyA2008-02156, AyA2011-24052; 3) the NASA/ IPAC Infrared Science Archive,
        which is operated by the Jet Propulsion Laboratory, California Institute of Technology, under contract with the National Aeronautics and Space Administration;
        4) the VizieR catalogue access tool, CDS, Strasbourg, France; and 4) Astropy, a community-developed core Python package for Astronomy
        \href{http://www.astropy.org}{Astropy Collaboration 2013}.
\end{acknowledgements}

\bibliographystyle{aa.bst}      
\bibliography{dzcha.bib}        

%

\end{document}